\documentclass[10pt,journal,compsoc]{IEEEtran}

\usepackage{nicefrac}
\usepackage{graphicx}
\usepackage{color}
\usepackage{amsmath, amsthm}
\usepackage{amsfonts}
\usepackage{enumerate}
\usepackage{algorithm, caption}
\usepackage[noend]{algpseudocode}
\usepackage{subcaption}
\usepackage{multirow}
\usepackage{booktabs}
\usepackage{dsfont}

\usepackage{comment}

\usepackage{tikz}
\usetikzlibrary{positioning}
\usepackage{pgfplots}
\usetikzlibrary{patterns}

\usepackage{tablefootnote}
\usepackage{subcaption}
\usepackage[section]{placeins}
\usepackage{hyperref}
\usepackage[open]{bookmark}
\usepackage{footmisc}

\newtheorem{theorem}{Theorem}
\newtheorem{note}{Note}
\newtheorem{remark}{Remark}
\newtheorem{problem}{Problem}

\newtheorem{lemma}[theorem]{Lemma}

\DeclareMathOperator*{\argmin}{arg\,min}  

\newcommand{\Tau}{\mathcal{T}}
\newcommand\numberthis{\addtocounter{equation}{1}\tag{\theequation}}
\newcommand{\Name}{\textsc{SsAG}}

\newenvironment{mybox}{\begin{tabular}{@{}l@{}}}{\end{tabular}}

\renewcommand{\vec}[1]{\mathbf{#1}}

\usepackage{enumitem}
\setlist[description]{leftmargin=0pt,labelindent=0pt}

\begin{document}

\title{\Name: Summarization and Sparsification of Attributed Graphs}

\author{Sarwan Ali, Muhammad Ahmad, Maham Anwer Beg, Imdad Ullah Khan, Safiullah Faizullah, Muhammad Asad Khan

\IEEEcompsocitemizethanks{
\IEEEcompsocthanksitem S. Ali is with Georgia State University, USA\protect\\
Email: sali85@student.gsu.edu
\IEEEcompsocthanksitem M. Ahmad is with Lahore University of Management Sciences, Pakistan\protect\\
Email: 17030056@lums.edu.pk
\IEEEcompsocthanksitem M. Beg is with Lahore University of Management Sciences, Pakistan\protect\\
Email: 14030016@lums.edu.pk
\IEEEcompsocthanksitem I. Khan is with Lahore University of Management Sciences, Pakistan\protect\\
Email: imdad.khan@lums.edu.pk
\IEEEcompsocthanksitem S. Faizullah is with The Islamic University, Madinah, KSA \protect\\
Email: safi@iu.edu.sa
\IEEEcompsocthanksitem M. Khan is with Hazara University, Mansehra, Pakistan\protect\\
Email: asadkhan@hu.edu.pk}
\thanks{Manuscript received April 19, 2005; revised August 26, 2015.}
}

\IEEEtitleabstractindextext{%
	
\begin{abstract}

We present \Name, an efficient and scalable lossy graph summarization method that retains the essential structure of the original graph. \Name{} computes a sparse representation (summary) of the input graph and also caters to graphs with node attributes. The summary of a graph $G$ is stored as a graph on supernodes (subsets of vertices of $G$), and a weighted superedge connects two supernodes. The proposed method constructs a summary graph on $k$ supernodes that minimize the reconstruction error (difference between the original graph and the graph reconstructed from the summary) and maximum homogeneity with respect to attributes. We construct the summary by iteratively merging a pair of nodes. We derive a closed-form expression to efficiently compute the reconstruction error after merging a pair and approximate this score in constant time. To reduce the search space for selecting the best pair for merging, we assign a weight to each supernode that closely quantifies the contribution of the node in the score of the pairs containing it. We choose the best pair for merging from a random sample of supernodes selected with probability proportional to their weights. A logarithmic-sized sample yields a comparable summary based on various quality measures with weighted sampling. We propose a sparsification step for the constructed summary to reduce the storage cost to a given target size with a marginal increase in reconstruction error. Empirical evaluation on several real-world graphs and comparison with state-of-the-art methods shows that \Name{} is up to $5\times$ faster and generates summaries of comparable quality. 

\end{abstract}

\begin{IEEEkeywords}
\texttt{graph summarization, attributed graphs, graph sparsification}
\end{IEEEkeywords} 
}

\maketitle

\section{Introduction}\label{sec:introduction}

Graphs analysis is a fundamental task in various research fields such as e-commerce, social networks analysis, bioinformatics, internet of things, etc.~\cite{CHEN2021102659,SONG2021102712}. Graphs with millions of nodes and billions of edges are ubiquitous in many applications. The magnitude of these graphs poses significant computational challenges for graph processing. A practical solution is to compress the graph into a summary that retains the essential structural information of the original graph. Processing and analyzing the summary is significantly faster and reduces the storage and communication overhead. It also makes visualization of very large graphs possible~\cite{ShenME06,ZhangBNCZ15}.

Graph summarization plays a pivotal role in drawing insights from a social or information network while preserving users' privacy~\cite{aggarwal2004condensation,HayMJTL10,Samarati01,Sweene02}. Summarization has been successfully applied to identify critical nodes for immunization to minimize the infection spread in the graph~\cite{PrakashVF12}. The summary of a graph helps efficiently estimate the combinatorial trace of the original graph, which is used to select nodes for immunization~\cite{Ahmad2020Trace,Ahmad2016AusDM,AhmadTSK17,TariqAKS17}. Moreover, summarization techniques help generate descriptors of graphs -  vectors representation of graphs for efficient graph analysis~\cite{AHMED2020169,Zohair2020Descriptors}. Summaries of attributed graphs are used for targeted publicity campaigns, nodes clustering, de-anonymization, and nodes attribute prediction~\cite{Khan20IPM,khan2015set}.


Given an undirected attributed graph  $G = (V,E,\mathcal{A})$, where $V$ and $E$ are sets of nodes and edges, respectively. $\mathcal{A}(v_i)$ is the attribute value for $v_i \in V$. Formally, $\mathcal{A}$ is a function mapping each node $v_i$ to one of the possible attribute values, i.e. $\mathcal{A}: V \mapsto \{a_1,a_2,\cdots,a_l\}$. 
For $k \in \mathbb{Z}^+$, a summary of $G$, $S=(V_S,E_S,\mathcal{A}_S)$ is a weighted graph on $k$ supernodes. $V_S = \{V_1,\ldots,V_k\}$ is a partition of $V$. Each $V_i$ has two associated integers, $n_i=|V_i|$ and $e_i= |\{(u,v): u,v \in V_i, (u,v) \in E \}|$. An edge $(V_i,V_j)\in E_S$ (called a superedge) has weight $e_{ij}$, where $e_{ij}$ is the number of edges in the bipartite subgraph induced between $V_i$ and $V_j$, i.e. $e_{ij}= |\{(u,v):u \in V_i, v \in V_j, (u,v) \in E \}|$. Each supernode $V_i$ has a $l$-dimensional feature vector that maintains the distribution of attribute values of nodes in $V_i$ i.e. $\mathcal{A}_S^{i}[p] = |\{ u_j \in V_i : \mathcal{A}(u_j) = a_p \}|, 1\leq p\le l$. 


Given a summary $S$ of a graph $G$, the original graph can be approximately reconstructed from $S$. The reconstructed graph $G'$ is represented by the expected adjacency matrix, ${A'}$ defined as 
\begin{equation}\label{eq:expectedAdjMatDef}
	{A'}(u,v) = \begin{cases} 0 & \text{ if } u = v\\ \nicefrac{e_i}{{n_i\choose 2}} & \text{ if } u,v\in V_i\\ \nicefrac{e_{ij}}{n_in_j} & \text{ if } u\in V_i, v\in V_j \end{cases}
\end{equation}
\begin{figure*}[h!]
	\centering
	\includegraphics[width=.9\textwidth,page = 5]{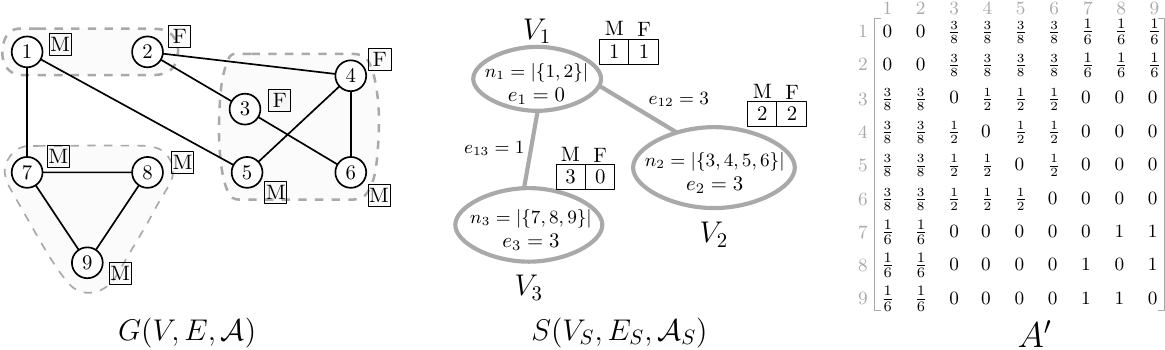}
	\caption{A graph $G$ with \textit{gender = male(M)/female(F)} attribute on nodes. $S$ on three supernodes is a summary of $G$. Each supernode $V_i$ has an attribute feature vector, $\mathcal{A}_S^{i}$, with the counts of males and females $V_i$. $RE_1(G|S) = 16.83$ and $purity(S) = 0.67$.The expected adjacency matrix $A'$ reconstructed from $S$ using Equation~\eqref{eq:expectedAdjMatDef}.}
	\label{fig:summaryExample}
\end{figure*}

The (unnormalized) $\ell_p$-reconstruction error, $RE_p$  of a summary $S$ of a graph $G$ is the $p^{th}$ norm of the error matrix ($A-A'$), where $A'$ is the expected adjacency matrix of $G'$. Note that $G'$ is approximate reconstruction of $G$ from $S$. Formally, $RE_p$ of a summary $S$ of graph $G$ is: 

\begin{equation}
	\resizebox{0.49\textwidth}{!}{
$RE_p(G|S) =  RE_p(A | {A'}) =  \big(\sum_{i=1}^{n} \sum_{j=1}^{n} |A (i,j) - A'(i,j)|^p\big)^{\nicefrac{1}{p}}$} \label{eq:RE}
\end{equation}
where $G|S$ denotes approximate reconstruction of $G$ from $S$ and $A|A'$ is the approximation of $A$ based on $S$. $RE_p$ quantifies the disagreements between $A$ and $A'$ constructed from $S$.



The attribute values for nodes are approximated from the feature vectors of supernodes in $S$. For a vertex $v_j \in V_i$, the probability that attribute value of $v_j$ is $a_p$ is $\nicefrac{\mathcal{A}_S^{i}[p]}{n_i}$.
A summary $S$ is homogeneous with respect to $\cal A$ if nodes in $V_i, i \le k$ have same attribute values. We use the purity of the partition $V_S = \{V_1,\ldots,V_k\}$ as a measure of the homogeneity of $S$. The purity of $S$, $purity(S)$ is computed as $purity(S) = \frac{1}{n}{\sum_{i=1}^k max(\mathcal{A}_S^{i})}$.  

The primary research objective of this work is to devise an efficient and scalable algorithm for the following dual optimization problem.
\begin{problem}\label{problem:1}
	Given a graph $G=(V,E,{\cal A})$ and a positive integer $k \leq n$, find a summary $S$ with $k$ supernodes such that $RE(G \vert S)$ is minimum and $purity(S)$ is maximum over all choices of $S$.
\end{problem}

As the number of possible summaries (vertex set partitions) is exponential, computing the {\em `best'} summary is challenging. A well-known method, \textsc{GraSS}~\cite{lefevre2010grass} uses an agglomerative approach to form a summary of the given graph. Initially, each node is considered a supernode, and in every iteration, two supernodes are merged until the desired number of supernodes are formed. Selecting a pair of nodes to merge and computing the error incurred after merging a pair are computationally expensive steps. At iteration $t$, let $n(t)$ be the number of supernodes in the summary, the possible number of pairs of supernodes is ${n(t)\choose 2}$. \textsc{GraSS} selects the best pair from a random sample of $O(n(t))$ pairs to reduce the search space. The selection and merging of the best pair take $O(n(t))$ time. The overall runtime to summarize the graph on $n$ nodes is $O(n^3)$. On the other hand, \textsc{s2l}~\cite{riondato2017graph} represents each node by an $n$-dimensional vector and applies vector-clustering to find supernodes. The runtime of \textsc{s2l} is $O(n^2t)$, where $t$ is the number of iterations. This runtime is still infeasible for large graphs. A recent and scalable method, SSumM~\cite{Lee2020Sparse}, summarizes a graph to minimize reconstruction error and keeps the summary size within a fixed size limit. However, none of these methods incorporate node attributes for summarization.



This paper proposes \Name, a lossy summarization approach that incorporates both the graph topology and node attributes. \Name{} constructs the summary by iteratively merging a pair of supernodes. We define a score function for pairs, a function of graph topology and attribute information, that quantify the reconstruction error and purity after merging a pair. For computational efficiency, we approximate this score using a closed-form expression and storing a constant number of extra variables at each node. We assign a weight to each node to estimate the contribution of the node in the score of pairs and select nodes with probability proportional to their weights. We choose the `best pair' for merging from a weighted random sample of nodes. Using weighted sampling and approximate score, \Name{} efficiently constructs a summary of comparable quality with logarithmic-sized samples. The overall runtime of \Name{} to compute a summary on $k$ supernodes is $O(n\log^2n)$. Furthermore, we use sparsification to reduce the storage size of the summary to a target budget with negligible impact on reconstruction error. 

The main contributions of this paper are as follows:
\begin{itemize}[leftmargin=8pt,labelindent=0pt]

    \item 
We define a score function for pairs of nodes to determine their goodness for merging that incorporates graph topology and node attribute values. This score can easily be approximated by storing a constant amount of extra information at each node.

    \item 
    We also define a weight for nodes that quantifies the contribution of a node in the score of pairs. A sample of $O(\log n)$ nodes selected with probability proportional to this weight yield results comparable to a linear-sized sample. With efficient score approximation and logarithmic-sized weighted sample, \Name{} takes $O(n\log^2n)$ time to construct a summary.
	
	\item 
	Experimental evaluation on several real-world networks shows that \Name{} significantly outperforms \textsc{GraSS} and \textsc{s2l} in terms of runtime and scalability to large graphs while producing comparable quality summaries.  
	\item 
	\Name{} produces homogeneous summaries with respect to attributes. Moreover, with sparsification, the summaries produced by \Name{} have low storage requirements comparable to the state-of-the-art approach of SSumM.
\end{itemize}


This manuscript is an extension of the earlier work~\cite{BegAZK18}, which describes the approximate scheme for the graph summarization based on the graph structure only. In this paper, we extend the idea of~\cite{BegAZK18} to incorporate the node attributes along with the graph topology to make a summary. We also provide detailed proofs and analyses of the summarization algorithm. Moreover, we present a sparsification approach to drop superedges to reduce the summary size while having a negligible impact on the reconstruction error. Lastly, we do an extensive evaluation of \Name{} based on various evaluation measures adopted in the literature.

The remaining paper is organized as follows. In Section ~\ref{section:related_work} we discuss previous work on graph summarization and related problems. We present our \Name{} along with its proof of correctness in Section~\ref{section:algorithm}. In Section~\ref{sec_performance_ana}, we give runtime analysis, the space complexity of the obtained summaries, and query computation based on a summary. The description of the dataset, baseline models, and evaluation metrics is given in Section~\ref{sec_experimental_setup}. We report the results of the empirical evaluation of \Name{} and its comparison with known methods in Section~\ref{section:experiments}. Finally, we conclude the paper in Section~\ref{section:conclusion}.

\section{Related work}\label{section:related_work}
Graph summarization and compression have been studied for a wide array of problems and have applications in diverse domains. It is widely used in clustering, classification, community detection, outlier identification, network immunization, etc.  In literature, there are two main types of graph summarization methods: \textit{lossless} and \textit{lossy}. In lossless summarization, the exact reconstruction is achieved by storing some extra information in the form of \textit{edge corrections} along with the summary~\cite{navlakha2008graph}. The edge corrections include the edges to be inserted (positive edge corrections) and edges to be deleted (negative edge corrections) from the reconstructed version of the graph. A scalable summarization approach summarizes sets of similar nodes that are found using locality sensitive hashing~\cite{khan2015set}. 

In lossy graph summarization, some detailed information is compromised to reduce the size and space complexity of the original graph. Reconstruction error~\cite{lefevre2010grass}, cutnorm error~\cite{riondato2017graph} and error in query answering are some of the widely used quality measures of a lossy summary. Reconstruction error is the norm of the error matrix (difference between the adjacency matrices of the original and the reconstructed version of the graph)~\cite{lefevre2010grass}. cutnorm error is defined as the maximum absolute difference between the sum of the weights of the edges between any two sets of vertices~\cite{riondato2017graph}. Similarly, accuracy in answers to various types of graph queries indicates the quality of the summary. For partitioning nodes into supernodes, an agglomerative approach is used in~\cite{lefevre2010grass} that greedily merges pairs of nodes to minimize the $l_1$-reconstruction error. This approach is very simple and achieves great summarization quality, but it does not scale to large graphs. Even after subsampling 
~\cite{lefevre2010grass}, the approach scales only to graphs of order of a few thousand. An efficient algorithm that uses a weighted sampling scheme was proposed in~\cite{BegAZK18} that can be applied to large-scale graphs.

Note that various types of graphs exist in different domains. Summarization on different types of graphs is applied to get useful results. In attributed graphs, nodes have certain associated attributes (properties)~\cite{Ali2020Predicting}. The lossless summarization of attributed graphs is described in~\cite{KhanNL17}, which identifies the sets of nodes having \textit{similar} neighborhood and attribute values for merging. Locality sensitive hashing is used to select nodes having similar neighborhood and attribute values. Moreover, to construct a summary with approximate homogeneous neighborhood information and attribute values using an entropy model is described in~\cite{liu2012approximate}. In addition to this, the summarization of attributed graphs based on a selected set of attributes is described in~\cite{TianHP08}. The work also presents an operation to allow users to \textit{drill down} to a larger summary for more details and \textit{roll up} to a concise summary with fewer details. Another line of work for attributed graphs finds a compact subgraph of the desired number of nodes having query attribute values~\cite{Khan20IPM}. A survey~\cite{You2013towards} discusses various summarization techniques for attributed graphs.

Compression of edge weighted graphs using locality sensitive hashing while preserving the edge weights is described~\cite{KhanDTNL17}. Furthermore, compression of node and edge weighted graphs is described such that the weights on the path between two nodes in the summary graph should be similar to that in the original graph~\cite{ZhouQT17}. The paper also aims to preserve more information related to nodes with high weights. Another closely related area is of influence graph in which nodes have influence over other nodes. The influence graph summarization takes into account the influence of nodes on other nodes in the summary graph~\cite{ShiTTL15}. 

Dynamic graphs, where nodes or edges evolve over time, are also prevalent these days. An approach discusses the summarization of a dynamic graph based on connectivity and communication among nodes~\cite{TsalouchidouMBB16}. The work creates summaries of the dynamic graph over a sliding window of fixed size.  MoSSo, a lossless approach, incrementally updates the summary in response to the deletions or additions of 
edges ~\cite{Ko2020Incremental}. A summarization framework that captures the dynamic nature of dynamic graphs is described in~\cite{QuLZJ16}.

Webgraph summarization improves the performance of search engines~\cite{boldi2004webgraph,SuelY01}. They are efficiently compressed by exploiting the link structure of the web~\cite{adler2001towards}. Permuting the nodes in a web graph such that {\em similar} nodes are placed together produces improved compression results~\cite{BoldiSV09}. Parallel methods are also devised to summarize massive web graphs spread over multiple machines~\cite{Shin2019LosslessAndLossy}.

Summarization of graph streams such that to approximately answer the queries on the graph stream is discussed in~\cite{0001CM16}. The real-time summarization of massive graph streams is done using the count-min sketch approach to preserve the structural information of the graph~\cite{KhanA17}.

Several methods use graph features as building blocks (vocabulary) for graph representation. \textit{VoG} (Vocabulary-based summarization of graphs) summarizes graphs based on the substructures like cliques, chains, stars, and bipartite cores~\cite{koutra2014vog}. Graph compression based on communities identified based on central nodes and hubs is studied in~\cite{LimKF14}. See ~\cite{liu2016graph} for a survey of graph summarization techniques.

Many summarization approaches result in very dense summaries. These summaries have $k$ supernodes, but the storage cost of superedges is very high; sometimes, it even surpasses that of the original graph. A recent scalable state-of-the-art approach, SSumM~\cite{Lee2020Sparse} summarizes the graph minimizing both the reconstruction error and summary density simultaneously. In this work, we utilized the notion of sparsification of SSumM to reduce the summary size to a target size. \cite{safavi2019personalized} proposed a framework to find a sparse summary of knowledge graphs preserving only the most relevant information. Authors in~\cite{papalampidi2020movie} propose a model for movie summarization by constructing a sparse movie graph by identifying important turning points in movies. The movie summarization model highlights different graph topologies for different movie genres.


\section{Proposed Solution}
\label{section:algorithm}

In this section, we present the details of the proposed solution, \Name{} for Problem~\ref{problem:1}. The goal is to construct a summary graph on $k$ supernodes that has the minimum reconstruction error $RE$ and maximum homogeneity of attribute values. After constructing the summary, we sparsify it to reduce the storage cost to the target size.

Given a graph $G= (V,E,\mathcal{A})$ on $n$ nodes,  an integer $k$ and target storage size, \Name{} produces a summary graph $S= (V_S,E_S, \mathcal{A}_S)$ on $k$ supernodes with storage size at most the given target. We give a general overview of \Name{} in Algorithm~\ref{basicAgglomerative}. $S$ is iteratively constructed in an agglomerative fashion. Initially, each node is a supernode, and in each iteration, a pair of supernodes is merged.  Denote by $S_{t}$ the summary after iteration $t$ with $(n-t)$ supernodes.  In iteration $t$, Algorithm~\ref{basicAgglomerative} selects a pair of nodes in $S_{t-1}$ for merging that results in the least $RE(G|S_t)$ and the maximum homogeneous merged supernode.

\begin{algorithm}[ht]
	\caption{: \textsc{Agglomerative-Summarization}($G=(V,E,\mathcal{A})$, $k$, \text{target size})}
	\label{basicAgglomerative} 
	\begin{algorithmic}[1]
		\State $S_0 \gets G$
		\State $t\gets 1$
		\While{$S_{t-1}$ has  more than $k$ nodes}
		\State Select the \emph{`best'} pair $(V_a,V_b)$ of (super)nodes in $S_{t-1}$ \Comment{{\bf T-1} and {\bf T-2}}
		\State Merge $V_a$ and $V_b$ in $S_{t-1}$ to get $S_{t}$ \Comment{{\bf T-3}}
		\EndWhile
		\State Sparsify $S_t$ to the target size \Comment{{\bf T-4}}
	\end{algorithmic}
\end{algorithm}

Let $n(t)$ be the number of supernodes in $S_t$ and $S_t^{(a,b)}$ be the graph obtained after merging $V_a$ and $V_b$ in iteration $t$. We identify the following key tasks in Algorithm \ref{basicAgglomerative}.

\begin{description}
\item[T-1: Efficient Score Computation of a Pair of Nodes:] While constructing a summary, we select a pair of supernodes for merging such that the resulting supernode has maximum homogeneity and the supergraph has the minimum reconstruction error.
For a given pair $(V_a,V_b)$ naive computation of only $RE(G \vert S_t^{(a,b)})$ from Equation~\eqref{eq:RE} takes $O(n^2)$ time. 
\item[T-2: Selection of the Best Pair of Nodes for Merging:]
By Lemma~\ref{lemma:scoreOfPair} the score of a pair can be estimated in constant time, however the search space is quadratic since there are $\binom{n(t)}{2}$ candidate pairs. This poses a significant hurdle to the scalability of Algorithm~\ref{basicAgglomerative} to large graphs.
\item[T-3: Efficient Merging of a Pair of Nodes:]
The next task is to merge the selected pair. For a summary with the adjacency list representation, naively merging a pair of nodes requires traversal of the whole graph in the worst case.

\item[T-4: Sparsification of Summary Graph:]
In many cases, the resulting summaries on $k$ supernodes are dense and even surpass the original graph in storage cost. The last task is to sparsify the summary to the given target size.

\end{description}

Next, we describe in detail how \Name{} accomplishes each of these tasks. Each supernode $V_a \in V_S$ stores $n_a$ and $e_a$. We store the weighted adjacency list of $S$, i.e. for edge $(V_a,V_b) \in E_S$ its weight $e_{ab}$ is stored at nodes $V_a$ and $V_b$ in the list. 
\subsection{{\bf T-1:} Efficient Score Computation of a Pair of Nodes}

We define a score for a pair that can be efficiently estimated with some extra book keeping. We need to select a pair of nodes $(V_a,V_b) \in \binom{V_{S_{t-1}}}{2}$ such that $RE(G\vert S_t^{(a,b)})$ is minimum. We denote ${score}_t^{{RE}} (a,b)$ as the score of $(V_a,V_b)$ based on graph topology, which quantifies the increase in $RE$ after merging the pair. Note that for fixed $S_{t-1}$, minimizing $RE(G\vert S_t^{(a,b)})$ is equivalent of maximizing $RE(G\vert S_{t-1})-RE(G\vert S_t^{(a,b)})$.

Each supernode $V_a$ has an attribute vector $\mathcal{A}_S^a$, which records the distribution of attribute values of nodes in $V_a$. Based on the attribute information, we compute the score of $V_a$ and $V_b$, $ {score}_t^{\mathcal{A}} (a,b)$ as $max(\mathcal{A}_S^a \oplus \mathcal{A}_S^b)$, where $\oplus$ denotes the element wise addition of  $\mathcal{A}_S^a \text{ and } \mathcal{A}_S^b$. Using a weight parameter $\alpha \in [0,1]$, we assign weight to the attribute and graph structure similarity and define the score of $(V_a,V_b)$ as

\begin{multline}
score_t(a,b) \;=\;  \alpha \times {score}_t^{{RE}} (a,b) + (1-\alpha) \times {score}_t^{\mathcal{A}} (a,b)  \\[.02in]
=  \alpha \big[ RE(G|S_{t-1}) - RE(G|S_t^{a,b}) \big] 
     + (1-\alpha) \text{max} \big( \mathcal{A}_S^a \oplus \mathcal{A}_S^b \big) 
\label{eq:scoreCombined}
\end{multline}

We normalize $RE(G\vert S_t^{(a,b)})$ and $\text{max} \big( \mathcal{A}_S^a \oplus \mathcal{A}_S^b)$ by $n^2$ and $(n_a + n_b)$, respectively to bring them in the same scale of $[0,1]$. To compute ${score}_t^{{RE}} (a,b)$, we need to efficiently evaluate the $RE$ of $S_t^{(a,b)}$. 

\begin{lemma}\label{lemma:scoreOfPair} 
	Given a summary $S_{t}$ with constant extra space per node, we can
\begin{enumerate}[label=(\roman*)]
		\item evaluate $score_t(a,b)$ in $O(n(t))$ time
		\item approximate $score_t(a,b)$ in constant time and space
	\end{enumerate}  
\end{lemma}
Note that score of a pair of nodes consists of two factors as given in Equation~\eqref{eq:scoreCombined}. We discuss the computation of each of the factors below, together the constitute a proof of Lemma\~ref{lemma:scoreOfPair}. 

 \begin{note}
At every supernode $V_a$ we store integers $n_a$ and $e_a$, the number of nodes in $V_a$ and the number of edges with both endpoints in $V_a$. Moreover, at $V_a$ we also store a real number $D_a =  \sum_{\substack{i=1\\i \neq a}}^{n(t)} \nicefrac{e_{ai}^2}{n_i}$, i.e. the sum of squares of weights of superedges incident on $V_a$. We can update $D_a$ in constant time after merging any two nodes $V_x,V_y \neq V_a$. After merging, we traverse neighbors list of $V_x$ and $V_y$ for $V_a$, subtract $\nicefrac{e_{xa}^2}{n_x}$ and $\nicefrac{e_{ya}^2}{n_y}$ from $D_a$ and add  $\nicefrac{(e_{x}+e_y)^2}{(n_x+n_y)}$ to $D_a$. 
 \end{note}

\subsubsection{$score_t^{RE}(a,b)$ computation} 

We derive a closed-form expression to efficiently compute $RE(G\vert S)$, which is the total error incurred in the estimation of $A$ from $S$ only. We first calculate the contribution of $V_a \in V_S$ in $RE(G\vert S)$ and then extend it to a general expression that sums the contribution of all the supernodes in $RE(G\vert S)$. Let $V_a = \{v_{a1},v_{a2},\cdots,v_{an_a}\}$, where $v_{aj} \in V, a\le k, j\le n_a$. $V_a$ contributes to all the entries/edges in $A'$, which have one or both the endpoints in $V_a$. We predict the presence of each of the possible internal edge (edge with both the endpoints in $V_a$) as $\nicefrac{e_a}{\binom{n_a}{2}}$. However, there are $e_a$ edges in $V_a$ and thus $2e_a$ corresponding entries in $A$ have value $1$. The error at these $2e_a$ entries in $A$ and $A'$ is $2e_a(1-\nicefrac{e_a}{\binom{n_a}{2}})$. The error at the remaining $2[\binom{n_a}{2} - e_a]$ positions is $2(\binom{n_a}{2}-e_a)(\nicefrac{e_a}{\binom{n_a}{2}})$. For a superedge $(V_a,V_b) \in E_S, V_a,V_b \in V_S, a,b\le k, a\ne b$, we predict the presence of an edge $(u,v), v\in V_a, u\in V_b, u,v\in V$ as $\nicefrac{e_{ab}}{n_an_b}$. Doing the similar calculation as above, the contribution of $V_a$ in the error at the entries $(u,v), u \in V_a,v\in V_b$ is $e_{ab}(1-\nicefrac{e_{ab}}{n_an_b}) + (n_an_b - e_{ab})(\nicefrac{e_{ab}}{n_an_b})$. 

Thus, the total error introduced by supernode $V_a$ is $2e_a(1-\nicefrac{e_a}{\binom{n_a}{2}})+2(\binom{n_a}{2}-e_a)(\nicefrac{e_a}{\binom{n_i}{2}})+e_{ab}(1-\nicefrac{e_{ab}}{n_an_b}) + (n_an_b - e_{ab})(\nicefrac{e_{ab}}{n_an_b})$. Simplifying this we get that the total error accumulated in reconstructing $A$ using summary $S$ with $k$ supernodes is       
\begin{equation} \label{eq:RE_closedForm}
\resizebox{0.495\textwidth}{!}{
$RE(G|S) =  RE(A | A') 
	= \sum\limits_{i=1}^{k}4e_i - \frac{4e_i^2}{\binom{n_i}{2}}  + \sum\limits_{i=1}^{k}\sum\limits_{\substack{j=1,j\neq i}}^{k}  2e_{ij} - \frac{2e_{ij}^2}{n_in_j}$
}
\end{equation}
Using Equation~\eqref{eq:RE_closedForm}, $score_t^{RE}(a,b)$ can be computed using Equation~\eqref{eq:scoreRE}.


{
\footnotesize
\begin{align*}
&score_t^{RE}(a,b) \;=\; RE(G|S_{t-1}) - RE(G|S_t^{a,b}) 
 \notag 
\\
	 =&~4e_a + 4e_b - \frac{4e_a^2}{\binom{n_a}{2}} - \frac{4e_b^2}{\binom{n_b}{2}} + \sum\limits_{\substack{i=1\\i \neq a,b}}^{k}4e_i -\frac{4e_i^2}{\binom{n_i}{2}}  \\
	 & + 2(2e_{ab} -\frac{2e_{ab}^2}{n_an_b}) +  \sum\limits_{\substack{i=1\\i \neq a,b}}^{k}  4e_{ai} - \frac{4e_{ai}^2}{n_an_i} +  \sum\limits_{\substack{i=1\\i \neq a,b}}^{k}  4e_{bi} \\ 
	 & - \frac{4e_{bi}^2}{n_bn_i}  + \sum\limits_{\substack{i,j=1\\i,j \neq a,b}}^{k} 2e_{ij} - \frac{2e_{ij}^2}{n_in_j} - 4\big(e_a+e_b+e_{ab}\big) \\
	 & + \frac{4\big(e_a+e_b+e_{ab}\big)^2}{\binom{n_a+n_b}{2}} - \sum\limits_{\substack{i=1\\i \neq a,b}}^{k}4e_i -\frac{4e_i^2}{\binom{n_i}{2}} \\ & - 4\sum\limits_{\substack{i=1\\i \neq a,b}}^{k} \Big(\big(e_{ai}+e_{bi}\big) - \frac{\big(e_{ai}+e_{bi}\big)^2}{\big(n_a+n_b\big)n_i}\Big) -\sum\limits_{\substack{i,j=1\\i,j \neq a,b}}^{k}2e_{ij} -\frac{2e_{ij}^2}{n_in_j} \\
	=& -\frac{4e_a^2}{\binom{n_a}{2}}  - \sum\limits_{\substack{i=1\\i \neq a}}^{n(t)} \frac{4e_{ai}^2}{n_an_i}  +\frac{4e_{ab}^2}{n_an_b}  - \frac{4e_b^2}{\binom{n_b}{2}}
 	\notag 
	- \sum\limits_{\substack{i=1\\i \neq b}}^{n(t)}\frac{4e_{bi}^2}{n_bn_i} \\
	& + \frac{4\big(e_a+e_b+e_{ab}\big)^2}{\binom{n_a+n_b}{2}}  
 	\notag 
	+\frac{4}{\big(n_a+n_b\big)}\sum\limits_{\substack{i=1\\i \neq a,b}}^{n(t)}\Big(\frac{e_{ai}^2}{n_i}+ \frac{e_{bi}^2}{n_i} + \frac{2e_{ai}e_{bi}}{n_i}\Big) 
 	\numberthis \label{eq:scoreRE}
\end{align*}
}

\begin{remark}\label{remark_compute_Score}
Given $n_a,n_b,e_a,e_b,e_{ab},D_a$ and $D_b$ we can compute all the terms of $score^{RE}_t(a,b)$ in constant time in Equation~\eqref{eq:scoreRE}, except for the last summation that requires a traversal of adjacency lists of $V_a$ and $V_b$. 
\end{remark}

The last summation in Equation~\eqref{eq:scoreRE}, $\sum_{\substack{i=1\\i \neq a,b}}^{n(t)}\nicefrac{2e_{ai}e_{bi}}{n_i}$, is the inner product of two $n(t)$ dimensional vectors $\vec{u}$ and $\vec{v}$, where the $i^{th}$ coordinate of $\vec{u}$ is $\nicefrac{e_{ai}}{\sqrt{n_i}}$ ($\vec{v}$ is similarly defined). It takes $O(n(t))$ space to store these vectors and $O(n(t))$ time to compute $score_t^{RE}(a,b)$. However, we can approximate $\left<\vec{u},\vec{v}\right> = \vec{u}\cdot \vec{v}$  using count-min sketch~\cite{Ali2019Detecting,CormodeM05}, which takes parameters $\epsilon \text{ and }\delta$. Let $w=\nicefrac{1}{\epsilon}$ and $d=\log \nicefrac{1}{\delta}$, a count-min sketch is represented by a $d\times w$ matrix. Using a randomly chosen hash function $h_i, 1\le i \le d$ such that $h_i:\{1 \dots n(t)\} \mapsto \{
1 \dots w\}$,  $j^{th}$ entry of $\vec{u}$ is hashed to $\big(j,h_i(j)\big)$ in the matrix. Let $\big<\widehat{ \vec{u},\vec{v} }\big>_i$ be the estimate using $h_i$, and set $\big<\widehat{ \vec{u},\vec{v} }\big> = \argmin_i \big<\widehat{ \vec{u},\vec{v} }\big>_i$.

\begin{theorem}\label{thm:CMSketch} (c.f~\cite{CormodeM05} Theorem 2) 
	For $0< \epsilon, \delta < 1$, let $\big<\widehat{ \vec{u},\vec{v} }\big>$ be the estimate for $\big<\vec{u},\vec{v}\big>$ using the count-min sketch. Then 
	
	\begin{itemize}
		\item $\big<\widehat{ \vec{u},\vec{v} }\big> \geq \big<\vec{u},\vec{v}\big>$\vskip.02in
		\item $Pr[\big<\widehat{ \vec{u},\vec{v} }\big> < \big<\vec{u},\vec{v}\big> + \epsilon ||\vec{u}||_1||\vec{v}||_1] \geq 1-\delta$
	\end{itemize}
	The computational cost of $\big<\widehat{ \vec{u},\vec{v} }\big>$ is $O(\nicefrac{1}{\epsilon}\log \nicefrac{1}{\delta})$ and updating the  sketch after a merge takes $O(\log\nicefrac{1}{\delta})$ time.
\end{theorem}

Hence, it takes constant time to closely approximate $score_t^{RE}(a,b)$ for a pair $(a,b)$ in $S_{t-1}$. Note that the bounds on time and space complexity, though constants are quite loose in practice.

\subsubsection{$score_t^{\mathcal{A}}(a,b)$ computation}$score_t^{\mathcal{A}}(a,b)$ computation for $V_a$ and $V_b$ is equivalent of element-wise addition of $\mathcal{A}_{S_{t-1}}^a$ and $\mathcal{A}_{S_{t-1}}^b$, which are the feature vectors containing the count of attribute values of nodes in $V_a$ and $V_b$, respectively. $score_t^{\mathcal{A}}(a,b)$ is the maximum value of the resultant feature vector. Note that the length of $\mathcal{A}_S^{a}$ is equal to $l$, the number of unique values of the attribute. So, the element-wise addition of $l$-dimensional feature vectors and then finding the maximum value of the resultant feature vector takes $O(l)$ time, where $l$ is a fixed constant (very small in practice).

\subsection{{\bf T-2:} Selection of the \emph{Best} Pair of Nodes for Merging} As in~\cite{lefevre2010grass}, we select the best pair out of a random sample; however,~\cite{lefevre2010grass} takes a sample of linear size, which is still computationally prohibitive for large graphs. We select nodes that are more likely to constitute high-scoring pairs. To this end, we define weights of nodes that essentially measure the contribution of a node to the score of the pair made up of this node. The weight, $w(a)$ of $V_a$ is given by

\begin{equation} \label{eq:weightOfNode} 
\resizebox{0.5\textwidth}{!}{
$w(a) = \begin{cases}
		\frac{-1}{f(a)} & \text{ if } f(a)\neq 0\\[.03in]
		0 & \text{ otherwise}
	\end{cases}
	\hskip.02in \text{\footnotesize{where}} \hskip.03in
	f(a) = -\frac{4e_a^2}{\binom{n_a}{2}}  - \sum\limits_{\substack{i=1\\i \neq a}}^{n(t)} \dfrac{4e_{ai}^2}{n_an_i}$
	}
\end{equation}

We sample nodes based on their weights, so nodes with higher weights are more likely to be sampled. This implies that pairs formed from these nodes will also have higher scores, i.e. will be of better quality. Using weighted sampling, a sample of size $O(\log n)$ outperforms a random sample of size $O(n)$. Recall that $n_a \text{ and }e_a,$ are stored at each supernode $V_a$. 
Using these variables, we update the weight of a given node in constant time. Let $W= \sum_{i=1}^{n(t)} w(i)$ denote the aggregate of node weights, the probability of selecting a node $V_a$ is $\nicefrac{w(a)}{W}$. In a given iteration, weighted sampling takes linear time but in the case of dynamic graphs, where nodes and edge weights change over time, weighted sampling is a challenging task. As in each iteration of summarization, we merge a pair of nodes and as a result of which, the weights of some other nodes also change. To address these issues, we design a data structure $\mathbb{T}$ for weighted sampling with following properties. 

\begin{lemma}\label{lemma:dataStructure}
	We implement $\mathbb{T}$ as a binary tree with following properties:
	\begin{enumerate}[label=(\roman*)]
		\item initially populating  $\mathbb{T}$ with $n$ nodes can be done in $O(n)$ time
		\item randomly sampling a node with probability proportional to its weight can be done in $O(\log n)$ time
		\item inserting, deleting, and updating weight of a node in $\mathbb{T}$ can be done in $O(\log n)$ time
	\end{enumerate}  
\end{lemma}

\textit{Data Structure for Sampling:} We implement $\mathbb{T}$ as a balanced binary tree and a leaf node in $\mathbb{T}$ represents a supernode in the graph. A leaf node also contains the weight and id of the supernode. Each internal node stores the sum of weights of its children and resultantly, the weight of root sums up to $\sum_{i=1}^{n(t)} w(i)$. Algorithm~\ref{alg:buildTree} presents the construction of $\mathbb{T}$ and we also show the structure of a tree node. The height of $\mathbb{T}$ is $\lceil \log n \rceil$ and it takes $O(n)$ time to construct the tree. We designed $\mathbb{T}$ independently, but later found out that it has been known to the statistics community since 1980~\cite{wong1980efficient}.


\begin{algorithm}[h!]
	\renewcommand{\thealgorithm}{}
	\floatname{algorithm}{Structure}
	\caption{Node}
	\begin{algorithmic}
		\State {\bf int} NodeId \Comment{contains the unique id of each vertex} \vskip 0.01in
		\State {\bf double} NodeWeight \Comment{stores weight of each node} \vskip 0.01in
		\State {\bf Node} *LeftChild \Comment{stores link to left child} \vskip 0.01in
		\State {\bf Node} *RightChild \Comment{stores link to right child} \vskip 0.01in
		\State {\bf Node} *ParentNode \Comment{stores pointer to parent node}
	\end{algorithmic}
\end{algorithm}

\begin{algorithm}[h!]
	\caption{\textsc{buildsamplingtree}($V$,$W$,$start$,$end$)} \label{alg:buildTree}
	\begin{algorithmic}[1]
		\If{$V[start] = V[end]$}
		\State leaf $\gets \Call{makenode}$ \vskip 0.01in
		\State leaf.NodeWeight $\gets W[start]$ \vskip 0.01in
		\State leaf.NodeId $\gets V[start]$ \vskip 0.01in
		\State \Return $leaf$
		\Else
		\State mid $\gets \frac{\text{start+end}}{2}$ \vskip 0.01in
		\State leftChild $\gets \Call{makenode}$ \vskip 0.01in
		\State leftChild $\gets \Call{buildsamplingtree}{V,W,start,mid}$ \vskip 0.01in
		\State rightChild $\gets \Call{makenode}$ \vskip 0.01in
		\State rightChild $\gets \Call{buildsamplingtree}{V,W,mid+1,end}$ \vskip 0.01in
		\State parent $\gets \Call{makenode}$ \vskip 0.01in
		\State parent.NodeWeight $\gets$ leftChild.NodeWeight + rightChild.NodeWeight \vskip 0.01in
		\State leftChild.ParentNode $\gets$ parent \vskip 0.01in
		\State rightChild.ParentNode $\gets$ parent \vskip 0.01in
		\State \Return $parent$
		\EndIf
	\end{algorithmic}
\end{algorithm}

A node $V_a$ is sampled form $\mathbb{T}$ with probability $\nicefrac{w(a)}{W}$ using Algorithm~\ref{alg:sample}, which takes as input the root node and a uniform random number $r\in[0,W]$. As $\mathbb{T}$ is a balanced binary tree, the sampling of a node takes $O(\log n)$ time (length of the path from the root to the node). To update the weight at a leaf, we begin at the leaf (using the stored pointer) and change the weight of that leaf. Following the parent pointers, we update the weights of internal nodes to the new sum of weights of children. Weight update of a leaf is equivalent to deleting that node form $\mathbb{T}$. Moreover, to efficiently insert a new node, we maintain a pointer to the first empty node. After merging a pair, the resulting supernode is inserted by updating the weight of the first empty leaf. The above information can be maintained and computed in $O(k)$ at a given time in $S_t$.

\begin{algorithm}[h!]
	\caption{: GetLeaf(Random Number $r$,$node$)}\label{alg:sample}
	\begin{algorithmic}[1]	
		\If{$\text{node.LeftChild} = \textsc{null}$  \textsc{ and }  $\text{node.RightChild}=\textsc{null}$}
		\State  \Return $\text{node.NodeId}$
		\EndIf	
		\If{$r < \text{node.LeftChild.NodeWeight}$} \vskip 0.01in
		\State \Return $\Call{getleaf}{r,\text{node.LeftChild}}$ 
		\Else
		\State \Return \Call{getleaf}{$r$ - node.LeftChild.NodeWeight, \hskip 0.001in node.RightChild}
		\EndIf	
	\end{algorithmic}
\end{algorithm}

\subsection{{\bf T-3:}  Merging of a Pair of Nodes}

We do the merging of a pair of nodes efficiently by doing extra bookkeeping per edge. As a preprocessing step, for each $(V_a,V_b)\in E_S$, in the adjacency list of $V_a$ at $V_b$, we store a pointer to the corresponding entry in the adjacency list of $V_b$.
\begin{lemma}\label{lem:merge}
	A pair $(V_a,V_b)$ of nodes in $S_{t-1}$, can be merged to get $S_{t}$ in time $O(deg(V_a) + deg(V_b))$.  
\end{lemma}

A pair $(V_a,V_b)$ of nodes in $S_{t-1}$, can be merged to get $S_{t}$ in time $O(deg(V_a) + deg(V_b))$. As we store $S$ in the adjacency list format, we need to iterate over neighbors of each $V_a$ and $V_b$ and record their information in a new list of the merged node. However, updating the adjacency information at each neighbor of $V_a$ and $V_b$ could lead to traversal of all the edges. As mentioned earlier, we maintain a pointer to each neighbor $x$ in the adjacency list of $V_a$. Using these pointers, we just need to traverse the lists of $V_a$ and $V_b$ and it takes constant time to update the merging information at each of the neighbors of $V_a$ and $V_b$. This preprocessing step takes $O(|E|)$ time at the initialization.

\subsection{{\bf T-4:} Sparsification of Summary Graph} 

We do sparsification (deletion of superedges) to reduce the storage size of $S$. The storage cost of $S$ with $k$ supernodes is given as in~\cite{Lee2020Sparse}:
\begin{equation}
    cost(S) = |E_S| \big( 2 \log_2k + \log_2(max(e_{ab})) \big) + n \log_2k
\end{equation}
Here, $E_S$ is the set of superedges in $S$, and $max(e_{ab})$ is the maximum weight of a superedge in $S$. Note that for any $S$ with $k$ supernodes, $|V| \log_2k$ will remain fixed and the storage cost depends on the number of superedges in $S$. Moreover, each superedge takes a constant number of bits and the change in the cost after dropping a superedge will remain constant; however, the change in reconstruction error varies for each superedge. The increase in reconstruction error after dropping a superedge $(V_a,V_b), a,b \le k, a\ne b$ is $2\big(\frac{e_{ab}}{n_an_b} -1\big)e_{ab}$. To do sparsification, we sort superedges based on the increase in reconstruction error and drop the desired number of edges that result in the minimum increase in reconstruction error. To reduce the storage size of $S$ to a certain target size, the number of superedges to be dropped are computed as $\frac{size(S) - \text{target size}}{2 \log_2k + \log_2(max(e_{ab}))}$. Sparsifying the summary takes $O(2k^2 \log k)$ time in the worst case as it sorts the superedges in $S$ by weight and then selects the desired number of superedges for deletion.


\section{\Name{} Performance Analysis and Summary Based Query Answering}\label{sec_performance_ana}
In this section, we analyze the time and space complexity of \Name{}. We also discuss how to compute answers to common graph analysis queries using only the summary. 

\subsection{Runtime Analysis of \Name} 
Algorithm~\ref{ourAlgo} is our main summarization algorithm that takes as input $G$, integer $k$ (target summary size), $s$ (sample size), $w$, $d$ ( $w=\nicefrac{1}{\epsilon}$ and $d = \log \nicefrac{1}{\delta}$ are parameters for count-min sketch) and $size_t$, the target storage cost of summary in bits. 


\begin{algorithm}[ht]
	\caption{: \Name($G=(V,E,\mathcal{A})$, $k$, $w$, $d$, $size_t$)}
	\label{ourAlgo} 
	\begin{algorithmic}[1]
	    \State $S\gets G$
		\State $\mathbb{T} \gets \Call{buildsamplingtree}{V,W,1,n}$ \Comment{Lemma~\ref{lemma:dataStructure}}
		\While{$|V_S| > k$}
		\State $ \text{pairs} \gets\Call{getsample}{\mathbb{T},s}$ 
		\State $ \text{pairScores} \gets\Call{computeapproxscore}{\text{pairs}, w,d}$ \Comment{Uses Equation~\eqref{eq:scoreCombined} and Theorem~\ref{thm:CMSketch}}
		\State  $(x,y) \gets\Call{max}{\text{pairScores}}$ \Comment{select the pair $(x,y)$ having maximum score}
		\State $\Call{merge}{x,y}$ \Comment{Lemma~\ref{lem:merge}}
		\For{each neighbor $u \in \{N(x) \cup N(y)\}$} \Comment{update weight of each neighbor of $x$ and $y$}
		\State $\Call{updateweight}{u,\mathbb{T}}$
		\EndFor
		\EndWhile
		\If{$\Call{size}{S}> size_t$}
		    \State $\text{delCount} \gets \frac{\Call{size}{S} - size_t}{2 \log_2k + \log_2(max(e_{ab}))}$ \Comment{count of superedges to be deleted from $S$}
		    \State $\Call{sparsify}{S,\text{delCount}}$
		\EndIf
	\end{algorithmic}
\end{algorithm}

As stated earlier, each node $V_a$ has a variable $D_a$, using which we can initialize weight array in $O(n)$ using Equation~\eqref{eq:weightOfNode}. By Lemma~\ref{lemma:dataStructure}, we populate $\mathbb{T}$ in $O(n)$ time and sample $s$ nodes from $\mathbb{T}$ in $O(s\log n)$ time (Line $3$). Score approximation of a pair of nodes takes constant time by Lemma~\ref{lemma:scoreOfPair} and it takes $O(\Delta)$ time to merge a pair by Lemma~\ref{lem:merge} (Line $6$), where $\Delta$ is the maximum degree in $G$. As updating and deleting the weights in $\mathbb{T}$ takes $O(\log n)$ time and the \textsc{while} loop makes $n-k+1$ iterations, the runtime to construct $S$ on $k$ nodes is $O((n-k+1)(s\log n + \Delta\log n)$. 
Typically, $k$ is a fraction of $n, k \in O(n)$ and we consider $s$ to be $O(\log n)$ or $O(\log^2 n)$. Generally, real world graphs are very sparse, ($\Delta$, the worst case upper bound, is constant). As mentioned earlier, sparsification takes $O(k^2 \log k)$ time, however, $k$ is a small fraction of $n$ and generally, $k^2 < n$. The overall complexity of \Name{} is $O(n\log^2 n)$ or $O(n\log^3n)$ for sample size of $O(\log n)$ or $O(\log^2 n)$, respectively.

\subsection{Space Complexity of Storing the Summary}
We give the space complexity of storing the summary $S$. Note that $S$ has $k$ supernodes and weights on both its nodes and edges. First, we need to store the mapping of $n$ nodes in $G$ to the $k$  supernodes in $S$. This is essentially a function that maps each node in $G$ to one of the possible $k$ locations. This mapping takes $O(n \log k)$ bits. For each supernode $V_i$, we maintain two extra variables $n_i$ and $e_i$. The cost to store these variables at nodes is $O(k\log n_i+ k\log \binom{n_i}{2}) = O(k\log n)$ bits. In addition to this, weights at superedges are also stored, which takes $O(k^2\log n)$ bits in total. So the overall space complexity of storing $S$ is $O(n\log k + k\log n +k^2\log n)$. 

\subsection{Summary-based query answering}\label{section:usingSummary}

A measure to assess the quality of a summary $S$ is the accuracy in answering queries about $G$ using only $S$. We give expressions to answer queries using $S$ efficiently. We consider widely used queries at different granularity levels~\cite{lefevre2010grass, riondato2017graph}. The node-level queries include degree, eigenvector-centrality, and attribute of a node. Degree and eigenvector-centrality query seek the number of edges incident on the query node and the node's relative importance in the graph. The attribute query answers the attribute value of the query node. We also discuss the adjacency query that asks whether an edge exists between the given pair of nodes. Finally, we compute the triangle density query, a graph level query, which gives the fraction of triangles in the graph. 




\par\smallskip\noindent{\textbf{Node Degree Query}} Given a node $v \in V_i$, degree of $v$ can be estimated using only $S$ as ${deg}'(v) = \sum_{j=1}^{n} {A}'(v,j)$. We can compute ${deg}'(v)$ in $O(k)$ time using the extra information stored at supernodes as ${deg}'(v) =\frac{1}{n_i}(2e_i + \sum_{j=1,j\neq i}^k e_{ij})$. 

\par\smallskip\noindent{\textbf{Node Eigenvector-Centrality Query}} Based on $S$, the estimate for eigenvector-centrality of a node $v$ is ${p}'(v)=\nicefrac{{deg}'(v)}{2| E |}$~\cite{lefevre2010grass}.  As we compute ${deg}'(v)$ in $O(k)$ time, we can compute eigenvector-centrality in $O(k)$ time. 

\par\smallskip\noindent{\textbf{Node Attribute Query}}
For a node $u$, the attribute query asks for the value of ${\mathcal{A}}(u)$. From $S$, we answer this query approximately as follows. For $u \in V_i$, the approximate answer ${\cal A}'(u)$ is $a_p$ with probability $\nicefrac{{\cal A}^i_S[p]}{n_i}$. 

\par\smallskip\noindent{\textbf{Adjacency Query}} 
Given two nodes $u,v \in V$, using $S$ the estimated answer to the query whether $(u,v)\in E$ is $A'(u,v)$. 

\par\smallskip\noindent{\textbf{Graph Triangle Density Query}} Let $t(G)$ be the number of triangles in $G$. $t(G)$ can be estimated from $S$ by counting $i)$ the expected number of triangles in each supernode, $ii)$ the expected number of triangles with one vertex in a supernode and the remaining two vertices in another supernode $iii)$ the expected number of triangles with the three vertices in three different supernodes. Let $\pi_{i} = \nicefrac{e_i}{\binom{n_i}{2}}$ and $\pi_{ij} = \nicefrac{e_{ij}}{n_in_j}$, then the estimate for $t(G)$, is  

\begin{equation*}
    \begin{aligned}
    {t}'(G) = & {} \sum\limits_{\substack{i=1}}^{k} \bigg[  
\binom{n_i}{3}\pi_{i}^3 + \sum\limits_{\substack{j=i+1}}^{k} \bigg(\pi_{ij}^2\bigg[\binom{n_i}{2}n_j\pi_{i} \\
& + \binom{n_j}{2}n_i\pi_{j}  \bigg] + \sum\limits_{\substack{l=j+1}}^{k}n_in_jn_l\pi_{ij}\pi_{jl}\pi_{il} \bigg)\bigg]
    \end{aligned}
\end{equation*}


Let $\Tau_i^a$, $\Tau_i^b$, $\Tau_i^c$, and $\Tau_i^d$ be the number of triangles made with three vertices inside $V_i$, with one vertex in $V_i$ and two in another $V_j$, with two vertices in $V_i$ and one in another $V_j$, and with one vertex in $V_i$ and two vertices in two distinct other supernodes $V_j$ and $V_l$, respectively. The number of each type of these triangles is given as: 

\begin{equation*}
    \begin{aligned}
    \Tau_i^a = &  {n_i\choose 3}\pi_{i}^3, \quad\Tau_i^b =  \sum\limits_{\substack{j=1\\j\neq i}}^{k}n_i {n_j\choose 2} \pi_{ij}^2 \pi_{j}, \\
    \Tau_i^c = & \sum\limits_{\substack{j=1\\j\neq i}}^{k} {n_i\choose 2} n_j \pi_{i}^2 \pi_{ij} \quad   \text{ and } \\
    \Tau_i^d  = & \sum\limits_{\substack{j=1\\j\neq i}}^{k} \sum\limits_{\substack{l=1\\j\neq i\\l\neq i,j}}^{k}  n_in_jn_l \pi_{ij} \pi_{jl} \pi_{il} 
    \end{aligned}
\end{equation*}


Keeping in view over-counting of each type of triangle, our estimate for $t(G)$ is  $$\sum\limits_{\substack{i=1}}^{k} \Tau_i^a + \nicefrac{\Tau_i^b}{2} + \nicefrac{\Tau_i^c}{2} + \nicefrac{\Tau_i^d}{3}.$$  Computing this estimate takes $O(k^2)$ time.

\section{Experimental Setup}\label{sec_experimental_setup}

This section describes the dataset, implementation details, evaluation metrics, and parameter values used for experimentation. We also describe the setup briefly for results comparison with baseline and SOTA methods. 
We perform our experiments on an Intel(R) Core i$5$ with $2.4$ GHz processor and $8$ GB memory using Java programming language and made the code available online for reproducibility~\footnote{\url{https://www.dropbox.com/sh/gh05kux3x04qp4r/AAAhVOf-RGdWFA2yv11V7Mzia?dl=0}}.

\subsection{Dataset Description and Statistics}

We perform experiments on real-world benchmark graphs. We treat all the graphs as undirected and unweighted. The order of the graphs ranges from a few thousand to a few million. The brief statistics and description of the graphs used in the experiments are given in Table~\ref{tbl:graph_stats}.

\begin{table}[h!]
		\centering
		\resizebox {0.5\textwidth} {!} {%
	\begin{tabular}{p{1.4cm} c c p{5.9cm}} 
		\toprule
		Name & $\vert V \vert$  & $\vert E \vert$  & Network Description \\ 
		\midrule  
	
		Caltech36~\cite{nr_dataset} & $769$ & $16,656$ & \begin{mybox} Facebook100 graph induced by Caltech users\\
		Nodes: Users, Edges: Friendships\\ Node attributes: Gender, High school status\end{mybox} \\ 
		\midrule
	
		\multirow{1}{2.1cm}{Political Blogs~\cite{kunegis2013konect}}  & $1,224$ & $16,718$ & \begin{mybox}Hyperlink Network \\ 
		Nodes: Political blogs, Edges: Hyperlinks \\ 
		Nodes attribute: Political view (liberal/conserv.) \end{mybox} \\  \midrule
	
		Facebook~\cite{snapnets} & $4,039$	 & 	$88,234$ &  \begin{mybox} Anonymized subgraph of Facebook\\ 
		Nodes: Facebook Users, Edges: Friendships\end{mybox}	\\  \midrule
	
		Email~\cite{snapnets} & $36,692$	 & 	$183,831$ & \begin{mybox}Online Social Network \\
		Nodes: Email addresses, Edges: Email exchange \end{mybox} \\  \midrule
	
		Stanford~\cite{snapnets} & $281,903$	 & 	$1,992,636$ & \begin{mybox}Web Network \\ 
		Nodes: Stanford webpages, Edges: Hyperlinks \end{mybox} \\   \midrule
	
		Amazon~\cite{snapnets} & $403,394$	 & 	$2,443,408$ & \begin{mybox}Co-purchasing network	 \\ 
		Nodes: Products, Edges: Co-purchased products \end{mybox} \\   \midrule
		
		YouTube~\cite{snapnets} & $1,157,828$	 & 	$2,987,624$ & \begin{mybox}
        Online Social Network \\
        Nodes: YouTube users, Edges: Friendships
		\end{mybox}	 \\ \midrule
		
		Skitter~\cite{snapnets} & $1,696,415$	 & 	$11,095,298$ & \begin{mybox}
		Internet Topology\\ 
		Nodes: Autonomous systems, Edges: Links
		\end{mybox}	  \\  \midrule
	
		Wiki-Talk~\cite{snapnets} & $2,394,385$	 & 	$4,659,565$ & \begin{mybox}
		Online Social Network\\
		Nodes: Wikipedia users, Edges: Edits on pages
		\end{mybox}	 \\
		
\bottomrule
\end{tabular} 
}
	\caption{Statistics and type of datasets used in experiments.}
	\label{tbl:graph_stats}
\end{table}

\subsection{Evaluation and Comparison Setup} 

We compare \Name{} with \textsc{GraSS}~\cite{lefevre2010grass} and \textsc{s2l}~\cite{riondato2017graph} in terms of computational efficiency and quality of summary. Furthermore, we make comparison with {SsumM}~\cite{Lee2020Sparse} to show the impact of sparsification. 

\subsubsection{\textsc{GraSS}~\cite{lefevre2010grass}}
\textsc{GraSS} generates a summary with the goal to  minimize the $l_1$-reconstruction error (RE). \textsc{GraSS} uses the agglomerative approach to generate a sequence of summaries on $i$ supernodes for $k\leq i \leq n$ of a graph on $n$ nodes. The total worst-case running time of 
\textsc{GraSS} is $O(n^3)$, thus it is feasible for graphs with a few thousand nodes. We compare the runtime and RE of \Name{} with \textsc{GraSS} only the Facebook dataset.

\subsubsection{\textsc{s2l}~\cite{riondato2017graph}}
\textsc{s2l} uses Euclidean-space clustering algorithm to generate a summary with $k$ supernodes. \textsc{s2l} minimizes the reconstruction error and evaluate summaries also on cutnorm error, storage cost, and accuracy in query answers. We compare our summaries with \textsc{s2l} on all these four measures. We also compare the runtimes of \textsc{s2l} and \Name{}.
In comparison with \textsc{s2l}, we report results for summary graphs with $k\in\{100,500\}$ supernodes for smaller datasets. For larger graphs (on more than $100000$ nodes) we make summaries on $k\in\{1000,2000\}$ supernodes. These numbers are chosen without any quality bias. We report results for \Name{} with the sample size parameter $s= 5 \log n(t)$, where $n(t)$ is the number of supernodes in the summary. Note that increasing $s$ improves the quality of the summary but increases the computational cost. To evaluate the impact of sample size on performance of \Name{} we also report its results with $s\in \{ \log n(t), 5\log n(t), \log^2 n(t) \}$. We compute approximate score using count-min sketch approach with width $w= 200\}$. 

\subsubsection{SSumM~\cite{Lee2020Sparse}}
Given the target summary size in bits (rather than a target number of supernodes), SSumM merges pairs of nodes that minimize the RE and result in a sparse summary. {SSumM} scores pairs based on both these objectives. The score function can be efficiently evaluated and is the building block of the greedy summarization strategy. Since the sparsity is an input parameter, we compare the RE in summaries of a fixed target size obtained using SSumM and \Name{}.

\subsubsection{Purity of summaries of attributed graphs} Since \Name{} incorporates attribute information along with the graph structure, we summarize widely-used attributed graph using \Name{}. We demonstrate the tradeoff between purity and reconstruction error using the user-set parameter $\alpha$. Note that in the current work, we only consider nominal attributes at nodes. \Name{} can readily be extended to incorporate numeric attributes.

\subsubsection{Scalability of \Name{}}
We show that \Name{} is scalable to larger real-world graphs on which the competitor methods can not be applied.  We ran \Name{} on graphs on more than $1$ million nodes and report RE and runtime for varying values of parameters.

\section{Results and Discussion}\label{section:experiments}
This section reports the results for the proposed model and its comparison with the baseline and SOTA methods. We first compare results of \Name{} with the baseline method, \textsc{GraSS}. Second, we give a detailed report on how \Name{} performs compared to \textsc{s2l} by the evaluation metrics mentioned above. We also evaluate the sparsification phase in \Name{} and demonstrate the effect of incorporating node attributes on reconstruction error. Finally, we demonstrate that \Name{} is scalable to very large graphs. In all experiments we set the (sketch depth) parameter $d= 2$ in Algorithm~\ref{ourAlgo}. 

\subsection{Comparison with \textsc{GraSS}}
\begin{figure}[h!]
    \centering
    \includegraphics[scale=0.7]{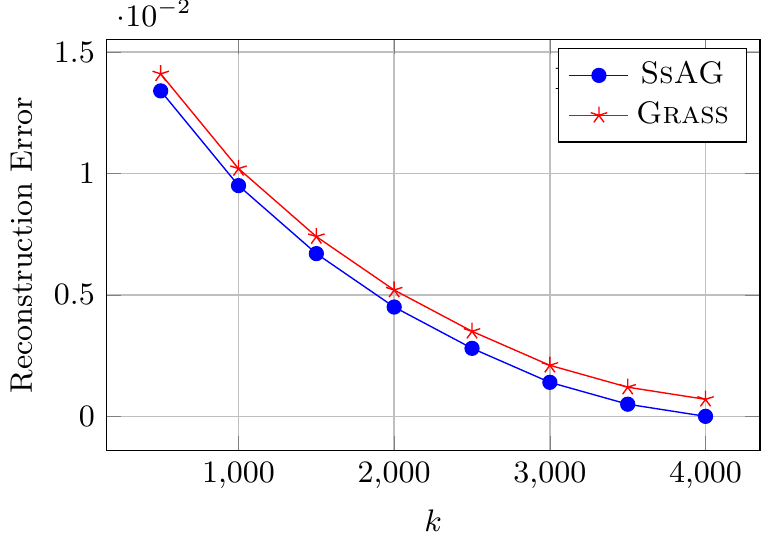}
    \caption{RE of \Name{} and \textsc{GraSS} for varying $k$ on Facebook graph. \textbf{Runtime to build a summary with $k=500$ of \Name{} and \textsc{GraSS} is 0.33 and 11.16 seconds, respectively.} }
	\label{fig:RE_Grass_cmp}
\end{figure}

We compare the normalized RE (scaled by $n\times n$, the number of entries in the adjacency matrix) with \textsc{GraSS} that only works for graphs with a few thousand nodes. We show comparison only on the Facebook graph ($4039$ vertices). Figure~\ref{fig:RE_Grass_cmp} depicts that reconstruction error increases with decreasing number of supernodes in the summary ($k$) for both \Name{} and \textsc{GraSS}, with \Name{} slightly outperforming \textsc{GraSS}. However, \Name{} is more than $33\times$ faster than \textsc{GraSS} as it took \textbf{0.33} seconds to build the summary while \textsc{GraSS} took \textbf{11.16} seconds. These results are for sample size $s  = 5\log n(t)$ in \Name. For other values of $s$, \Name{} exhibits the expected trend; with increasing sample size, RE decreases at the cost of running time (see Section~\ref{secSampleImpact}).
%
%
%
\begin{table}[b!]
	\centering
	\renewcommand\tabcolsep{2pt}
	\resizebox{1.02\linewidth}{!}{
	\begin{tabular}{p{0.25cm} c c c c c c c c c}
	\toprule
		$G$ & $k$ & Method & $w$ & RE & \begin{mybox}Cutnorm\\Error\end{mybox} & \begin{mybox}Storage\\Cost \\(KB)\end{mybox} & \begin{mybox}Avg. \\Degree \\Error
\end{mybox}& \begin{mybox}Triangle \\Density \\Error
\end{mybox} & \begin{mybox}Time \\ (sec)\end{mybox}  \\
\midrule
		
\multirow{8}{*}{\rotatebox[origin=c]{90}{Facebook}} & \multirow{4}{*}{100} &  \multirow{2}{*}{\Name{}} &  
200 & 1.65E-2 & 3.64E-3 & 7.86 & 16.74$\pm$21 & \textbf{-0.52} & \textbf{0.19} \\ 
& & & $\Box$ &  1.62E-2 & 4.10E-3 & 8.65 & 17.18$\pm$20 & -0.49 & 0.32\\ 
& & \textsc{s2l} & \textsc{n/a} & \textbf{1.06E-2} & \textbf{3.05E-3} & \textbf{6.57} & \textbf{9.89}$\pm$\textbf{12} & -0.30 & 1.45\\ 
\cmidrule(lr){2-10}

 & \multicolumn{3}{c}{\%Improv. of $\Box$ from \textsc{s2l}} & -52.83 & -34.43 & -31.66 & -73.71 & -63.33 & 77.93 \\
\cmidrule(lr){2-10}
 & \multirow{4}{*}{500} & \multirow{2}{*}{\Name{}} & 
 
200 & 1.35E-2 & 3.60E-3 & 62.5 & 11.99$\pm$15 & -0.30 & \textbf{0.22} \\ 
& & & $\Box$ & 1.32E-2 & \textbf{2.48E-3} & 81.00 & 11.22$\pm$12 & -0.28 & 0.37\\ 
& & \textsc{s2l} & \textsc{n/a} & \textbf{8.61E-3} & 2.85E-3 & \textbf{39.71} & \textbf{7.21}$\pm$\textbf{8} & \textbf{-0.32} & 4.68\\

\cmidrule(lr){2-10} 
 & \multicolumn{3}{c}{\%Improv. of $\Box$ from \textsc{s2l}} & -53.31 & 12.98 & -103.98 & -55.62 & 12.50 & 92.09 \\
\midrule 

\multirow{8}{*}{\rotatebox[origin=c]{90}{Email}} & \multirow{4}{*}{100} & \multirow{2}{*}{\Name{}} & 
200 & 5.28E-4 & 1.91E-4 & 44.34 & 6.79$\pm$25 & \textbf{-0.80} & \textbf{1.96} \\ 
& & & $\Box$ & 5.26E-4 & \textbf{1.39E-4} & 46.45 & 6.31$\pm$14 & -0.76 & 2.57\\ 
& & \textsc{s2l} & \textsc{n/a} & \textbf{5.00E-4} & 2.40E-4  & \textbf{37.99} & \textbf{5.70}$\pm$\textbf{16} & -0.77 & 45.94\\ 

\cmidrule(lr){2-10} 
 & \multicolumn{3}{c}{\%Improv. of $\Box$ from \textsc{s2l}} & -5.20 & 42.08 & -22.27 & -10.70 & 1.30 & 94.41 \\
\cmidrule(lr){2-10} 
 & \multirow{4}{*}{500} & \multirow{2}{*}{\Name{}} & 
200 & 5.18E-4 & \textbf{1.38E-4} & 153.36 & 5.71$\pm$19 & -0.65 & \textbf{2.79} \\ 
& & & $\Box$ & 4.89E-4 & 2.21E-4 & 158.74 & 5.15$\pm$10 & -0.63 & 4.1\\ 
& & \textsc{s2l} & \textsc{n/a} & \textbf{4.49E-4} & 1.71E-4  & \textbf{112.34} & \textbf{4.79}$\pm$\textbf{12} & \textbf{-0.73} & 55.4\\ 

\cmidrule(lr){2-10} 
 & \multicolumn{3}{c}{\%Improv. of $\Box$ from \textsc{s2l}} & -8.91 & -29.24 & -41.30 & -7.52 & 13.70 & 92.60 \\
\midrule 

\multirow{8}{*}{\rotatebox[origin=c]{90}{Stanford}} & \multirow{4}{*}{1000} & \multirow{2}{*}{\Name{}} & 
200 & 7.16E-5 & \textbf{3.49E-5} & 860.11 & 7.69$\pm$41 & \textbf{-0.68} & \textbf{89.43} \\ 
& & & $\Box$ & 7.10E-5 & 3.77E-5 & 876.4 & 8.06$\pm$42 & -0.67 & 108.78\\ 
& & \textsc{s2l} & \textsc{n/a} & \textbf{5.37E-5} & 6.30E-5 & \textbf{374.04} & \textbf{5.11}$\pm$\textbf{36} & -0.26 & 305.19\\ 

\cmidrule(lr){2-10} 
 & \multicolumn{3}{c}{\%Improv. of $\Box$ from \textsc{s2l}} & -32.22 & 40.1 & -134.31 & -57.73 & -157.69 & 64.36 \\
\cmidrule(lr){2-10} 
 & \multirow{4}{*}{2000} & \multirow{2}{*}{\Name{}} & 
200 & 6.91E-5 & \textbf{3.16E-5} & 1417.6 & 7.38$\pm$40 & \textbf{-0.64} & \textbf{80.6} \\ 
& & & $\Box$ & 6.87E-5 & 5.18E-5 & 1462.5 & 7.60$\pm$38 & -0.63 & 109.36\\ 
& & \textsc{s2l} & \textsc{n/a} & \textbf{4.65E-5} & 6.80E-5  & \textbf{449.29} & \textbf{4.06}$\pm$\textbf{10} & -0.23 & 425.95\\

\cmidrule(lr){2-10} 
 & \multicolumn{3}{c}{\%Improv. of $\Box$ from \textsc{s2l}} & -47.74 & 23.82 & -225.51 & -87.19 & -173.91 & 74.33 \\
\midrule 
\multirow{8}{*}{\rotatebox[origin=c]{90}{Amazon}} & \multirow{4}{*}{1000} & \multirow{2}{*}{\Name{}} & 
200 & 5.98E-5 & 3.04E-5 & 537.66 & 5.63$\pm$13 & \textbf{-1.00} & \textbf{496.17} \\ 
& & & $\Box$ & 5.97E-5 & \textbf{2.89E-5} & 537.29 & 5.64$\pm$13 & \textbf{-1.00} & 632.35\\ 
& & \textsc{s2l} & \textsc{n/a} & \textbf{5.91E-5} & 4.20E-5  & \textbf{509.50} & \textbf{5.37}$\pm$\textbf{11} & -0.96 & 993.00\\ 

\cmidrule(lr){2-10} 
 & \multicolumn{3}{c}{\%Improv. of $\Box$ from \textsc{s2l}} & -1.02 & 31.19 & -5.45 & -5.03 & -4.17 & 36.32 \\
\cmidrule(lr){2-10} 
 & \multirow{4}{*}{2000} & \multirow{2}{*}{\Name{}} & 
200 & 5.96E-5 & \textbf{2.84E-5} & 648.02 & 5.57$\pm$12 & -0.99 & \textbf{793.57} \\ 
& & & $\Box$ & 5.95E-5 & 2.88E-5 & 638.55 & 5.60$\pm$13 & -0.99 & 1096.01\\ 
& & \textsc{s2l} & \textsc{n/a} & \textbf{5.81E-5} & 3.80E-5  & \textbf{584.29} & \textbf{5.13}$\pm$\textbf{9} & \textbf{-0.92} & 1115.78\\ 
\cmidrule(lr){2-10}
 & \multicolumn{3}{c}{\%Improv. of $\Box$ from \textsc{s2l}} & -2.41 & 24.21 & -9.29 & -9.16 & -7.61 & 1.77\\ 
	\bottomrule
\end{tabular}
}
 \caption{Comparison of \Name{} with \textsc{s2l} on different evaluation metrics. Observe that \Name{} is better by Cutnorm error than \textsc{s2l} for all but the smallest dataset (Facebook with $k=100$). Moreover, with slight degradation in performance \Name{} significantly outperforms \textsc{s2l} by runtime. Sample size $s  = 5\log n(t)$ and scores of pairs are computed exactly ($w =\Box$) and approximated with count-min sketch width $w =200$. Best values are shown in bold. We also show \% improvement of \Name{} ($w = \Box$) from \textsc{s2l} using Equation~\eqref{eq_percent_increase}.}
\label{tbl:comparisonRE} 
\end{table}

\pgfplotsset{title style={at={(0.7,0.8)}}}
\pgfplotsset{compat=1.5}
\noindent
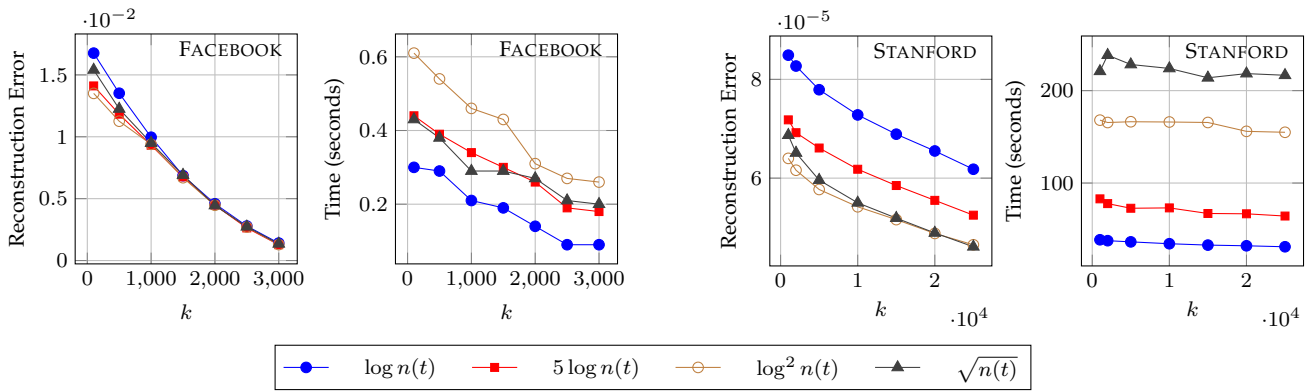
\begin{figure*}[h!]
	\centering\footnotesize
	\begin{tikzpicture}
	\begin{axis}[title={\textsc{Facebook}},
	height=0.522\columnwidth, width=0.51\columnwidth, grid=major,
	ylabel={ Reconstruction Error},
	xlabel={$k$},
	]
	\addplot+[ blue, mark size=2pt, mark =*, mark options={blue}] table[x={x},y={RE_log_n}, col sep=comma]{Figures_data/Facebook_RE_sample_size_impact.csv};
	\addplot+[ red, mark size=1.5pt, mark=square*, mark options={red}] table[x={x},y={RE_5_log_n}, col sep=comma]{Figures_data/Facebook_RE_sample_size_impact.csv};
	\addplot+[ brown, mark size=2pt, mark =o, mark options={brown}] table[x={x},y={RE_log2_n}, col sep=comma]{Figures_data/Facebook_RE_sample_size_impact.csv};
	\addplot+[ darkgray, mark size=2.5pt, mark=triangle*, mark options={darkgray}] table[x={x},y={RE_sqrt_n}, col sep=comma]{Figures_data/Facebook_RE_sample_size_impact.csv};
	\end{axis}
	\end{tikzpicture}
	\begin{tikzpicture}
	\begin{axis}[title={\textsc{Facebook}},
	height=0.522\columnwidth, width=0.51\columnwidth, grid=major,
	ylabel={Time (seconds)},
	xlabel={$k$},
	legend style={
		column sep=3ex,
	},
	legend entries={$\log n$,$5\log n$,$\log^2n$,$\sqrt{n}$},
	legend to name=commonlegend,
	]
	\addplot+[ blue, mark size=2pt, mark =*, mark options={blue}] table[x={x},y={time_log_n}, col sep=comma]{Figures_data/Facebook_RE_sample_size_impact.csv};
	\addplot+[ red, mark size=1.5pt, mark=square*, mark options={red}] table[x={x},y={time_5_log_n}, col sep=comma]{Figures_data/Facebook_RE_sample_size_impact.csv};
	\addplot+[ brown, mark size=2pt, mark =o, mark options={brown}] table[x={x},y={time_log2_n}, col sep=comma]{Figures_data/Facebook_RE_sample_size_impact.csv};
	\addplot+[ darkgray, mark size=2.5pt, mark=triangle*, mark options={darkgray}] table[x={x},y={time_sqrt_n}, col sep=comma]{Figures_data/Facebook_RE_sample_size_impact.csv};
	\end{axis}
	\end{tikzpicture}
	\hskip.4in
	\begin{tikzpicture}
	\begin{axis}[title={\textsc{Stanford}},
	height=0.522\columnwidth, width=0.51\columnwidth, grid=major,
	ylabel={ Reconstruction Error},
	xlabel={$k$},
	]
	\addplot+[ blue, mark size=2pt, mark =*, mark options={blue}] table[x={x},y={RE_log_n}, col sep=comma]{Figures_data/Stanford_RE_sample_size_impact.csv};
	\addplot+[ red, mark size=1.5pt, mark=square*, mark options={red}] table[x={x},y={RE_5_log_n}, col sep=comma]{Figures_data/Stanford_RE_sample_size_impact.csv};
	\addplot+[ brown, mark size=2pt, mark =o, mark options={brown}] table[x={x},y={RE_log2_n}, col sep=comma]{Figures_data/Stanford_RE_sample_size_impact.csv};
	\addplot+[ darkgray, mark size=2.5pt, mark=triangle*, mark options={darkgray}] table[x={x},y={RE_sqrt_n}, col sep=comma]{Figures_data/Stanford_RE_sample_size_impact.csv};
	\end{axis}
	\end{tikzpicture}
	\begin{tikzpicture}
	\begin{axis}[title={\textsc{Stanford}},
	height=0.522\columnwidth, width=0.51\columnwidth, grid=major,
	ylabel={ Time (seconds)},
	xlabel={$k$},
	legend columns=4,
	legend style={
		column sep=3ex,
	},
	legend entries={$\log n(t)$,$5\log  n(t)$,$\log^2n(t)$,$\sqrt{n(t)}$},
	legend to name=commonlegend,
	]
	\addplot+[ blue, mark size=2pt, mark =*, mark options={blue}] table[x={x},y={time_log_n}, col sep=comma]{Figures_data/Stanford_RE_sample_size_impact.csv};
	\addplot+[ red, mark size=1.5pt, mark=square*, mark options={red}] table[x={x},y={time_5_log_n}, col sep=comma]{Figures_data/Stanford_RE_sample_size_impact.csv};
	\addplot+[ brown, mark size=2pt, mark =o, mark options={brown}] table[x={x},y={time_log2_n}, col sep=comma]{Figures_data/Stanford_RE_sample_size_impact.csv};
	\addplot+[ darkgray, mark size=2.5pt, mark=triangle*, mark options={darkgray}] table[x={x},y={time_sqrt_n}, col sep=comma]{Figures_data/Stanford_RE_sample_size_impact.csv};
	\end{axis}
	\end{tikzpicture}
	\vskip.1in
	\ref{commonlegend}
	\caption{Impact of sample size on RE (left column) and runtime (right column). We take sample size $s  \in \{ \log n(t),5\log n(t),\log^2n(t),\sqrt{n(t)} \}$, where $n(t)$ is the number of supernodes in the summary. Increasing $s$ results in significant increase in runtime, however, there is relatively less reduction in RE.}
	\label{fig:RE_sample_size_impact}
\end{figure*}

\subsection{Comparison with \textsc{s2l}}
In this section, we report the results of an extensive comparison of \Name{} and \textsc{s2l} based on RE, cutnorm Error, Storage Cost (KB), Average Degree Error, Triangle Density Error, and Computation Time. Table~\ref{tbl:comparisonRE} shows the results for $4$ datasets, namely Facebook, Email, Stanford, and Amazon. To show the approximation quality of score computation, we report results for exact score computation using Equation~\eqref{eq:scoreRE} and Theorem~\ref{thm:CMSketch}. We used $w = \nicefrac{1}{\epsilon} = 200$, other values show a similar trend. Results show that, especially for large graphs, the count-min sketch closely approximates the exact score of a pair for merging. The rows corresponding to \Name{} with exact score computation are listed with $w=\Box$. These results are for sample size $s  = 5\log n(t)$ in \Name. We demonstrate the effect of different sample sizes on RE and runtime in Section~\ref{secSampleImpact}. We also provide $\%$ improvement for \Name{} (with $w = \Box$) from \textsc{s2l} for each metric using the expression:
\begin{equation}\label{eq_percent_increase}
    \text{\% increase} \;\;=\;\; \dfrac{\text{\Name{}  value $-$ \textsc{s2l} value}}{\text{\textsc{s2l} value}} \times 100
\end{equation}

\par\smallskip\noindent{\bf Runtime}
Computational efficiency is the most salient feature and prominent achievement of \Name{}. Table~\ref{tbl:comparisonRE} shows that runtime of \Name{} is significantly smaller in all settings compared to \textsc{s2l} but results in slightly larger reconstruction error. As expected, estimating pairs' scores with sketch takes less time at the cost of a negligible increase in RE. Observe that on Email dataset, \Name{} is $\sim17\times$ faster than \textsc{s2l}, while RE of \Name{} is only $0.0026\%$ worse than that of \textsc{s2l}. Similarly, for Stanford dataset, \Name{} is $\sim5\times$ more efficient while the difference in error is only $2\times 10^{-5}$. 

\par\smallskip\noindent{\bf  Cutnorm Error}
The cutnorm-error in the summaries produced by \Name{} is less than those by \textsc{s2l} for all graphs  except for the small Facebook graph (when $k=100$). However, cutnorm error for \Name{} is better for Facebook dataset when we have $k=500$. We observe the trend that \Name{} substantially outperform \textsc{s2l} for small $k$. We achieved up to $42\%$ improvement in cutnorm error compared to \textsc{s2l} (for Email dataset when $w = \Box$).

\par\smallskip\noindent{\bf Storage Cost} 
Table~\ref{tbl:comparisonRE} gives the space consumption of summaries produced by \Name{} and \textsc{s2l}. The summaries produced by \Name{} have higher storage cost compared to those made by \textsc{s2l}. Considering the storage size of the summary in terms of the percentage of the size of the original graph, the difference in summary sizes of the two approaches is not very significant. For the Amazon dataset, the summaries on $1000$ supernodes by \textsc{s2l} and \Name{} take $4.58\%$ and $4.83\%$ space of the original graph, respectively. For Stanford dataset, the summaries with $k=2000$ by \textsc{s2l} and \Name{} take $5\%$ and $15\%$ storage cost of the original graph. Note that the high memory cost is due to the dense summary graph (many superedges with small weights). To reduce their storage sizes, we sparsify the summaries that cause a slight increase in reconstruction error (see Section \ref{subsec:sparsificationResults}).

\par\smallskip\noindent{\bf Accuracy in Query Answers}
In Table~\ref{tbl:comparisonRE}, we report the mean and standard deviation of errors in node degrees estimated from summaries and errors in answers to the (relative) triangle density query in the graphs. The results follow the expected trend that the accuracy of query answers improves as $k$ increases for both \Name{} and \textsc{s2l}. Also the accuracy increases for large $w$ in \Name{}.

\subsection{Impact of Sample Size on RE and Runtime}\label{secSampleImpact} 
We show the impact of sample size on RE and time taken to compute the summary. We evaluate \Name{} with $s \in \{\log n(t), 5\log n(t), \sqrt{n(t)}, \log^2 n(t) \}$ and $w=200$ to show the impact of $s$ on the quality of the summary and runtime. In Figure~\ref{fig:RE_sample_size_impact}, it is evident that the error decreases with the increase in $s$, however, the benefit in quality is not proportional to the increase in computational cost. We show the results for Facebook and Stanford datasets; other datasets show the same trend regarding the quality of summary and computational time with varying $s$.

\subsection{Sparsification and Comparison with SSumM} \label{subsec:sparsificationResults}
\begin{figure}[h!]
	\centering
	\begin{subfigure}{.31\linewidth}
		\centering
		\includegraphics[scale = 0.6] {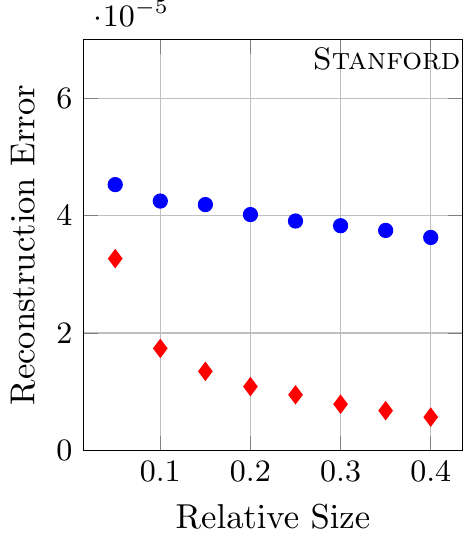}
	\end{subfigure}%
	\hskip.07in 
	\begin{subfigure}{.31\linewidth}
		\centering
		\includegraphics[scale = 0.6] {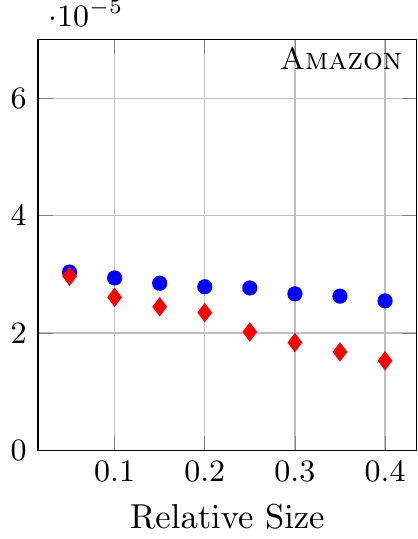}
	\end{subfigure}%
	\hskip.01in 
	\begin{subfigure}{.31\linewidth}
		\centering
		\includegraphics[scale = 0.6] {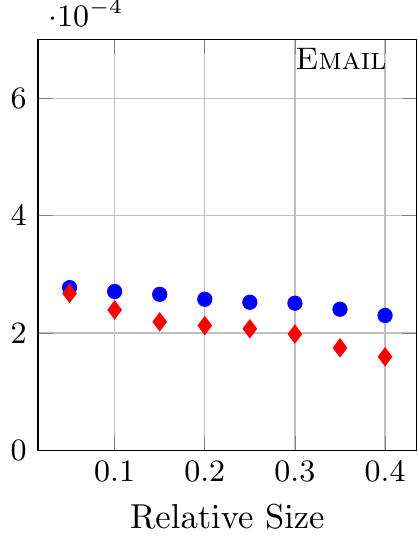}
	\end{subfigure}%
	\\[.05in]
	\begin{subfigure}{.25\textwidth}
		\centering
		\includegraphics[scale = 0.7] {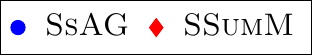}
	\end{subfigure}%
	\caption{RE of \Name{} and SSumM at varying relative sizes ($\frac{\text{summary size}}{\text{original graph size}}$) of summary graphs on three datasets. The results are reported on a scale of $10^{-5}$ and $10^{-4}$ and the max. absolute difference in RE of \Name{} and SSumM (Stanford dataset) at relative size $=0.4$ is $0.003\%$.}
	\label{fig:sparsification_cmp}
\end{figure}

In this section, we report RE of summaries of a fixed size obtained by \Name{} and SSumM. Figure~\ref{fig:sparsification_cmp} shows RE of summaries with varying relative sizes ($\frac{\text{summary size}}{\text{original graph size}}$) of three graphs. 
\pgfplotsset{title style={at={(0.30,0.03)}}}
\pgfplotsset{compat=1.5}
\noindent
\begin{figure*}[h!]
	\centering\footnotesize
	\begin{tikzpicture}
	\begin{axis}[title={Caltech36(Status)},
	axis y line* = left, 
	height=0.602\columnwidth, width=0.602\columnwidth, grid=major,
	legend style={
		column sep=3ex,
	},
	xlabel={{\small $k$}},
	ylabel={{\small Reconstruction Error}}]
	\addplot+[blue, mark size=1.5pt, mark=square*, mark options={blue},mark repeat=2] table[x={x},y={a_1_RE}, col sep=comma]{Figures_data/Caltech36_purity_attribute_1_tikz.csv};
	\addplot+[red, mark size=1.5pt, mark=square*, mark options={red},mark repeat=2] table[x={x},y={a_0.5_RE}, col sep=comma]{Figures_data/Caltech36_purity_attribute_1_tikz.csv};
	\addplot+[darkgray, mark=otimes,mark size=1.5pt, mark options={darkgray},mark repeat=2] table[x={x},y={a_0_RE}, col sep=comma]{Figures_data/Caltech36_purity_attribute_1_tikz.csv};
	\end{axis}
	\begin{axis}[
	height=0.602\columnwidth, width=0.602\columnwidth,
	ymin=0, ymax=1.05,
	hide x axis,
	hide y axis]
	\addplot+[blue, dashed, mark size=1.5pt, mark=square*, mark options={blue},mark repeat=2] table[x={x},y={a_1_Atr}, col sep=comma]{Figures_data/Caltech36_purity_attribute_1_tikz.csv};
	\addplot+[red, dashed, mark size=1.5pt, mark=square*, mark options={red},mark repeat=2] table[x={x},y={a_0.5_Atr}, col sep=comma]{Figures_data/Caltech36_purity_attribute_1_tikz.csv};
	\addplot+[darkgray, dashed, mark=otimes,mark size=1.5pt, mark options={darkgray},mark repeat=2] table[x={x},y={a_0_Atr}, col sep=comma]{Figures_data/Caltech36_purity_attribute_1_tikz.csv};
	\end{axis}
	\begin{axis}[
	ymin=0, ymax=1.05,
	hide x axis,
	axis y line*=right,
	height=0.602\columnwidth, width=0.602\columnwidth,
	]
	\end{axis}
	\end{tikzpicture}
	\begin{tikzpicture}
	\begin{axis}[title={Caltech36(Gender)},
	axis y line* = left, 
	height=0.602\columnwidth, width=0.602\columnwidth, grid=major,
	legend style={
		column sep=3ex,
	},
	xlabel={{\small $k$}},
	]
	\addplot+[blue, mark size=1.5pt, mark=square*, mark options={blue},mark repeat=2] table[x={x},y={a_1_RE}, col sep=comma]{Figures_data/Caltech36_purity_attribute_2_tikz.csv};
	\addplot+[red, mark size=1.5pt, mark=square*, mark options={red},mark repeat=2] table[x={x},y={a_0.5_RE}, col sep=comma]{Figures_data/Caltech36_purity_attribute_2_tikz.csv};
	\addplot+[darkgray, mark=otimes,mark size=1.5pt, mark options={darkgray},mark repeat=2] table[x={x},y={a_0_RE}, col sep=comma]{Figures_data/Caltech36_purity_attribute_2_tikz.csv};
	\end{axis}
	\begin{axis}[
	height=0.602\columnwidth, width=0.602\columnwidth,
	ymin=0, ymax=1.05,
	hide x axis,
	hide y axis]
	\addplot+[blue, dashed, mark size=1.5pt, mark=square*, mark options={blue},mark repeat=2] table[x={x},y={a_1_Atr}, col sep=comma]{Figures_data/Caltech36_purity_attribute_2_tikz.csv};
	\addplot+[red, dashed, mark size=1.5pt, mark=square*, mark options={red},mark repeat=2] table[x={x},y={a_0.5_Atr}, col sep=comma]{Figures_data/Caltech36_purity_attribute_2_tikz.csv};
	\addplot+[darkgray, dashed, mark=otimes,mark size=1.5pt, mark options={darkgray},mark repeat=2] table[x={x},y={a_0_Atr}, col sep=comma]{Figures_data/Caltech36_purity_attribute_2_tikz.csv};
	\end{axis}
	\begin{axis}[
	ymin=0, ymax=1.05,
	hide x axis,
	axis y line*=right,
	height=0.602\columnwidth, width=0.602\columnwidth,
	]
	\end{axis}
	\end{tikzpicture}
	\begin{tikzpicture}
	\begin{axis}[title={Political Blogs},
	axis y line* = left, 
	height=0.602\columnwidth, width=0.602\columnwidth, grid=major,
	legend style={at={(0.0,-0.2)},anchor=north},
	legend style={
		column sep=3ex,
	},
	xlabel={{\small $k$}},
	]
	\addplot+[blue, mark size=1.5pt, mark=square*, mark options={blue},mark repeat=1] table[x={x},y={a_1_RE}, col sep=comma]{Figures_data/PoliticalBlogs_RE_purity_attribute_1_tikz.csv};\label{re_alpha_1}
	\addplot+[red, mark size=1.5pt, mark=square*, mark options={red},mark repeat=2] table[x={x},y={a_0.5_RE}, col sep=comma]{Figures_data/PoliticalBlogs_RE_purity_attribute_1_tikz.csv};\label{re_alpha_0.5}
	\addplot+[darkgray, mark=otimes,mark size=1.5pt, mark options={darkgray},mark repeat=2] table[x={x},y={a_0_RE}, col sep=comma]{Figures_data/PoliticalBlogs_RE_purity_attribute_1_tikz.csv};\label{re_alpha_0}
	\end{axis}
	\begin{axis}[
	height=0.602\columnwidth, width=0.602\columnwidth,
	ymin=0, ymax=1.05,
	legend columns=6,
	legend style={at={(0,-0.3)},anchor=north, column sep=2ex},
	legend entries={$Purity_{\alpha =1}$,$Purity_{\alpha =0.5}$,$Purity_{\alpha =0}$},
	legend to name=commonlegend,
	hide x axis,
	hide y axis]
	\addlegendimage{/pgfplots/refstyle=re_alpha_1}\addlegendentry{$RE_{\alpha =1}$}
	\addlegendimage{/pgfplots/refstyle=re_alpha_0.5}\addlegendentry{$RE_{\alpha =0.5}$}
	\addlegendimage{/pgfplots/refstyle=re_alpha_0}\addlegendentry{$RE_{\alpha =0}$}
	\addplot+[blue, dashed, mark size=1.5pt, mark=square*, mark options={blue},mark repeat=2] table[x={x},y={a_1_Atr}, col sep=comma]{Figures_data/PoliticalBlogs_RE_purity_attribute_1_tikz.csv};
	\addlegendentry{$Purity_{\alpha =1}$}
	\addplot+[red, dashed, mark size=1.5pt, mark=square*, mark options={red},mark repeat=2] table[x={x},y={a_0.5_Atr}, col sep=comma]{Figures_data/PoliticalBlogs_RE_purity_attribute_1_tikz.csv};
	\addlegendentry{$Purity_{\alpha =0.5}$}
	\addplot+[darkgray, dashed, mark=otimes,mark size=1.5pt, mark options={darkgray},mark repeat=2] table[x={x},y={a_0_Atr}, col sep=comma]{Figures_data/PoliticalBlogs_RE_purity_attribute_1_tikz.csv};
	\addlegendentry{$Purity_{\alpha =0}$}
	\end{axis}
	\begin{axis}[
	ymin=0, ymax=1.05,
	hide x axis,
	axis y line*=right,
	height=0.602\columnwidth, width=0.602\columnwidth,
	ylabel={{\small Purity}},
	]
	\end{axis}
	\end{tikzpicture}
	\\ 
	\ref{commonlegend}
	\caption{Summarization of attributed graphs Caltech36 and Politcal Blogs, where nodes in Caltech36 have two attributes (status and gender) while Political blogs has one attribute (affiliation). The Left $y$-axis shows reconstruction error while right $y$-axis shows the purity of summary with $k$ supernodes. Results are reported with $\alpha \in \{0,0.5,1\}$; $\alpha = 0.5$ gives equal weight to the graph topology and node attributes.} 
	\label{fig:attributedSummPolBlogs}
\end{figure*}
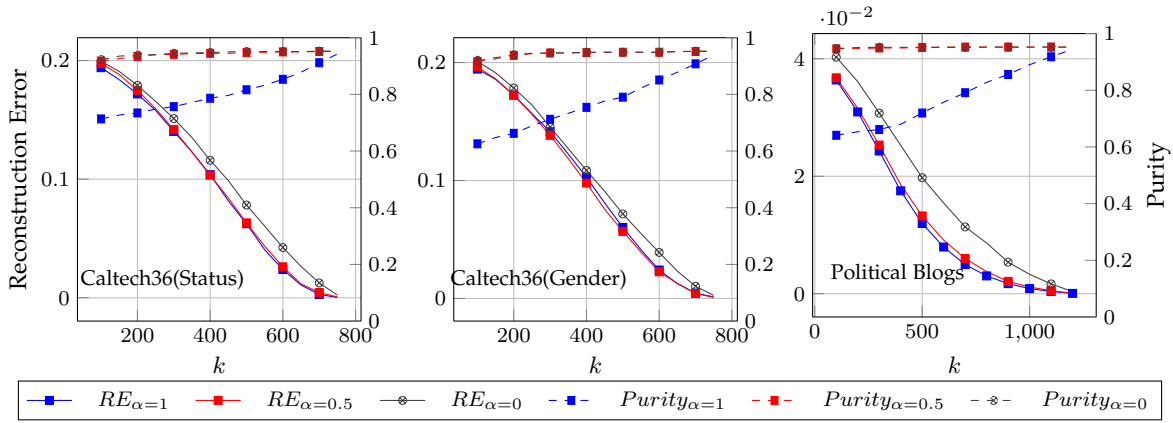
For the Stanford graph, after sparsification, the summary size is reduced to $5\%$ of the original graph size and achieves RE of $4.5\times10^{-5}$. This is less than the RE of the summary of the corresponding size by \textsc{s2l}. Compared to SSumM, summaries obtained using \Name{} are similar in quality. The difference in RE of summaries is not very significant (the $y$-axis represents values at $10^{-5}$ and $10^{-4}$ scale). For the Stanford graph, the maximum absolute difference in RE of \Name{} and SSumM at a relative size of $0.4$ is $0.003\%$. On other graphs, the difference in RE is significantly smaller.

\subsection{Summarization of Attributed Graphs: RE vs Purity}
 
\Name{} also incorporates node attributes for summarization along with the graph structure. We report purity of summaries obtained by \Name{} using $\alpha \in [0,1]$ as a  weight parameter for the graph topology and attribute information ($\alpha = 1$ means summarization only based on graph structure while $\alpha = 0$ mean summarization only based on attributes). We report the results in Figure~\ref{fig:attributedSummPolBlogs} for $\alpha \in \{0,0.5,1\}$ that show that with increasing weight of attributes, \Name{} yields summaries with high purity and a minimal increase in RE.

  
  

\begin{table}[h!]
	\centering
	\renewcommand\tabcolsep{3pt}
	\resizebox{1.01\linewidth}{!}{
	\begin{tabular}{c c c c c c c c c c c}	
	\toprule
		&&\multicolumn{2}{c}{Skitter     }&\multicolumn{2}{c}{Wiki-Talk	}&\multicolumn{2}{c}{YouTube}\\
		&&\multicolumn{2}{c}{$|V|$= 1,696,415   }&\multicolumn{2}{c}{$|V|$ = 2,394,385}&\multicolumn{2}{c}{$|V|$ =1,157,828} \\	
		&&\multicolumn{2}{c}{$|E|$= 11,095,298  }&\multicolumn{2}{c}{$|E|$ = 4,659,565	}&\multicolumn{2}{c}{$|E|$ = 2,987,624	}\\
		
		\cmidrule(lr){3-4}	\cmidrule(lr){5-6} \cmidrule(lr){7-8}			

		$k\times10^3$&w&$RE$&Time(s)&$RE$&Time(s)&$RE$&Time(s)\\ \midrule
		
		\multirow{4}{*}{10}&50& $2.29$E-6&521.43 &$1.56$E-6&311.10& $3.34$E-6&207.38\\ 
		&100& $2.23$E-6&516.03& $1.51$E-6&328.19& $3.22$E-6&222.22\\ 
		&200& $2.19$E-6&559.91& $1.46$E-6&363.37& $3.11$E-6&251.94\\ 
		&$\Box$& $2.14$E-6&649.82& $1.44$E-6&319.95& $3.09$E-6&242.58\\ 
		\midrule
		\multirow{4}{*}{50}
		&50& $2.11$E-6&481.40& $1.23$E-6&285.89& $2.50$E-6&184.67\\ 
		&100& $1.96$E-6&480.85& $1.21$E-6&299.98& $2.38$E-6&195.85\\ 
		&200& $1.88$E-6&524.94& $1.20$E-6&329.39& $2.37$E-6&215.87\\ 
		&$\Box$& $1.97$E-6&591.35& $1.20$E-6&273.24& $2.36$E-6&199.48\\
		\midrule
		\multirow{4}{*}{100}
		&50& $1.91$E-6&436.84& $1.09$E-6&266.15& $1.97$E6&160.11\\ 
		&100& $1.70$E-6&445.27& $1.09$E-6&276.32& $1.94$E-6&167.67\\ 
		&200& $1.66$E-6&486.88& $1.09$E-6&303.44& $1.94$E-6&183.64\\ 
		&$\Box$& $1.66$E-6&535.02& $1.09$E-6&248.73& $1.94$E-6&164.80\\ 
		\midrule
		\multirow{4}{*}{250}
		&50& $1.38$E-6&332.27& $9.07$E-7&223.65& $1.30$E-6&103.25\\ 
		&100& $1.24$E-6&350.47& $9.05$E-7&232.39& $1.30$E-6&107.79\\ 
		&200& $1.23$E-6&376.89& $9.03$E-7&256.05& $1.30$E-6&118.73\\ 
		&$\Box$& $1.23$E-6&392.58& $9.03$E-7&203.93& $1.30$E-6&98.70\\ 
		\bottomrule
	\end{tabular}
	}
	\caption{Reconstruction error and runtime (seconds) to compute summaries of large graphs using \Name{}. Approximate score is computed using count-min sketch width $w\in \{50,100,200\}$ and exact score computation ($w=\Box$).} 
	\label{tbl:resultsLargeGraph}
\end{table}
\subsection{Scalability of \Name} 
To demonstrate the scalability of \Name{}, we run \Name{} on graphs on more than $1$ million nodes. Table~\ref{tbl:resultsLargeGraph} reports quality and runtime only for \Name{} on large graphs, since \textsc{s2l} is not applicable to graphs at this scale. \Name{} can compute the summaries of large datasets in a few minutes while \textsc{s2l} fails to compute the summaries on these datasets even after running for a whole day. These results are for sample size $s  = 5\log n(t)$. Note that increasing the value of $w$ does not improve the quality summaries of large graphs. 

\section{Conclusion}\label{section:conclusion}
We propose \Name, a sampling-based efficient approximate method that considers both the graph structure and node attribute information for summarization. We build the summary by iteratively merging pairs of nodes. The pair is selected based on the score quantifying the reconstruction error resulting after merging it. We approximate the score in constant time with theoretical guarantees using the closed-form expression. Experimental results on several benchmark datasets show that our technique is comparable in quality and performs better than competitor methods in efficiency and runtime. Moreover, \Name{} also incorporates attributed graphs and produces highly homogeneous summaries. Furthermore, our sparsification method greatly reduces the size of the summary graph without significantly impacting the reconstruction error. In the future, we hope to evaluate \Name{} on edge attributed graphs for experiments along with more real-world graphs having node attributes. 

\bibliographystyle{IEEEtran}
\bibliography{GraphSummarization_Journal}

\end{document}